\documentclass[12pt]{article}
\usepackage{ifpdf}
\usepackage{pgf}
\pgfdeclareimage[width=15cm]{1a}{fig1aa}
\pgfdeclareimage[width=15cm]{1b}{fig1bb}
\pgfdeclareimage[width=15cm]{2a}{fig2aa}
\pgfdeclareimage[width=15cm]{2b}{fig2bb}
\usepackage{caption}
\usepackage{amsmath}
\usepackage{amstext}
\usepackage{amsthm}
\usepackage{amssymb}
\usepackage{amsopn}
\usepackage{mathrsfs} 
\usepackage{dsfont} 
\usepackage[bottom]{footmisc}
\usepackage{indentfirst}
\usepackage{upref}
\usepackage{cite}
\usepackage{units}
\usepackage{fancybox}
\usepackage[lflt]{floatflt}
\numberwithin{equation}{section}
\renewcommand{\baselinestretch}{1.4}
\DeclareMathOperator{\tr}{Tr}
\DeclareMathOperator{\ee}{e}
\newcommand{\chib}{\chi_{\text{B}}}
\newcommand{\etan}{\eta_{\text{N}}}
\newcommand{\dd}{\mathrm{d}}
\newcommand{\p}{\partial}
\newcommand{\h}{\widehat}
\newcommand{\w}{\widetilde}
\newcommand{\ov}{\overline}
\begin{document}
\begin{titlepage} 
\renewcommand{\baselinestretch}{1.1}
\small\normalsize
\begin{flushright}
MZ-TH/08-04
\end{flushright}

\vspace{0.1cm}

\begin{center}   

{\LARGE \textsc{Background Independence and \\[2mm]Asymptotic Safety in \\[6mm]Conformally Reduced Gravity}}

\vspace{1.4cm}
{\large M.~Reuter and H.~Weyer}\\

\vspace{0.7cm}
\noindent
\textit{Institute of Physics, University of Mainz\\
Staudingerweg 7, D--55099 Mainz, Germany}\\

\end{center}

\vspace*{0.6cm}
\begin{abstract}
We analyze the conceptual role of background independence in the application of the effective average action to quantum gravity. Insisting on a background independent renormalization group (RG) flow the coarse graining operation must be defined in terms of an unspecified variable metric since no rigid metric of a fixed background spacetime is available. This leads to an extra field dependence in the functional RG equation and a significantly different RG flow in comparison to the standard flow equation with a rigid metric in the mode cutoff. The background independent RG flow can possess a non-Gaussian fixed point, for instance, even though the corresponding standard one does not. We demonstrate the importance of this universal, essentially kinematical effect by computing the RG flow of Quantum Einstein Gravity in the ``conformally reduced'' Einstein--Hilbert approximation which discards all degrees of freedom contained in the metric except the conformal one. Without the extra field dependence the resulting RG flow is that of a simple $\phi^4$-theory. Including it one obtains a flow with exactly the same qualitative properties as in the full Einstein--Hilbert truncation. In particular it possesses the non-Gaussian fixed point which is necessary for asymptotic safety.
\end{abstract}
\end{titlepage}
%
%
%
%
%
%
\section{Introduction}\label{s1}
Finding a logically consistent and predictive quantum theory of gravity continues to be one of the most challenging open problems in theoretical physics. Even though the recent years have seen considerable progress in loop quantum gravity, string theory, and asymptotic safety, to mention just three approaches \cite{kiefer}, it seems that certain essential ingredients of a satisfactory microscopic theory are still missing or only poorly understood. In any of these approaches one typically encounters problems which are conceptually very difficult and deep, and at the same time highly complex from the calculational point of view.
On the conceptual side, the most severe problem is perhaps the issue of background independence \cite{A,R,T}. Already classically General Relativity is distinguished from all other physical theories in that it does not only tell us how physical processes take place in a given spacetime but also describes the dynamics of spacetime itself. Many problems one encounters when searching for a quantum theory of gravity can be traced back to this crucial property of General Relativity, namely that it dynamically generates the ``arena'' in which all physics is going to take place. In particular, the mediator of the gravitational interaction, the metric or closely related fields, defines the proper length or mass scale of all dimensionful physical quantities.

In the following we investigate a particular aspect of background (in)dependence which is particularly important in the context of asymptotic safety \cite{wein}, \cite{mr} - \cite{livrev}.
In this approach gravity is described by a quantum field theory of the metric tensor which is renormalized nonperturbatively at a non-Gaussian renormalization group (RG) fixed point. In order to implement this idea one has to pick a concrete RG framework. In principle many choices are possible here; they differ by the generating functionals they employ, in the way field configurations get ``integrated out'' along the RG flow, and, related to that, the interpretation of the corresponding RG scale, henceforth denoted $k$. In theories on flat spacetime there exist implementations of the Wilsonian RG, the effective average action \cite{avact,ym,avactrev,ymrev} for instance, which have the special property that the mass scale $k$ has a ``quasi--physical'' meaning in the following sense: The basic functional RG equation (FRGE) describes the $k$-dependence of a family of effective actions $\left\{ \Gamma_k,~0 \leq k < \infty \right\}$ each of which defines an effective field theory valid near the scale $k$.
This is to mean that in a single--scale problem involving typical momenta $p$ the optimal effective field theory description\footnote{Here the notion of an effective field theory is to be understood in the sense that a tree level evaluation of $\Gamma_k$ is sufficient for taking the quantum effects which are dominant near $k$ into account (provided the fluctuations are small enough).} is provided by $\Gamma_k$ with $k=p$.

Going over to quantum gravity it is not clear a priori how one could introduce an RG scale with a comparable physical meaning. 
The problem is that if $k$ is to have the status of a physical momentum it must be a \textit{proper} rather than merely a \textit{coordinate} momentum. However, proper momenta, distances, or other dimensionful quantities require a metric for their definition, and if the metric is dynamical it is not clear with respect to which metric $k$ should be ``proper''. 
Proceeding naively, the average action of gravity would be a functional $\Gamma_k [g_{\mu \nu}]$ which, besides $k$, depends on a single argument $g_{\mu \nu}$. More precisely, $\Gamma_k [\,\cdot\,]$, for every fixed value of $k$, is a map from the space of metrics into the reals. This implies that from the point of view of $\Gamma_k [\,\cdot\,]$ with $k$ fixed all metrics have an equal status so that $k$ cannot be ``proper'' with respect to any particular one of them. This is a direct consequence of background independence. It entails that the naive implementation of the average action idea, leading to a family of functionals $\left\{ \Gamma_k \right\}$ which depend only on one metric argument, cannot be labeled by an RG scale with the above ``quasi--physical'' interpretation.

The situation can also be described as follows. In standard quantum field theory on flat space the RG flow is a universal object in the sense that it does not depend on any specific field configuration. From the perspective of the running coupling constants $\left\{ g_n (k) \right\}$ the functional $\Gamma_k$ has the character of a generating function.
In a theory with field(s) $\phi$ and basis functionals (monomials) $\left\{ P_n [\phi] \right\}$ one usually expands 
$\Gamma_k [\phi] = \sum_n g_n (k) \, P_n [\phi]$.
By definition, $g_n (k)$ is the component of $\Gamma_k$ in the direction of the ``basis vector'' $P_n$. The role of $\phi$ is merely that of a dummy parameter which is needed to distinguish the basis vectors. It is obvious, therefore, that the RG trajectories 
$k \mapsto \left\{ g_n (k) \right\}$ bear no relationship to any specific field configuration $\phi$. If we use the same formalism for the metric in quantum gravity then $k$ cannot be a ``proper'' momentum therefore.

The actual effective average action for gravity constructed in \cite{mr} achieves the desired ``quasi--physical'' status of $k$ by using the background technique. The idea is to fix an arbitrary background metric $\overline{g}_{\mu \nu}$, quantize the (not necessarily small) metric fluctuations $h_{\mu \nu}$ nonperturbatively in this background, and finally adjust $\overline{g}_{\mu \nu}$ in such a way that the expectation value of the fluctuation vanishes: 
$\overline{h}_{\mu \nu} \equiv \langle h_{\mu \nu} \rangle =0$.
In this way the background gets fixed dynamically. The advantage of this procedure is that the actual quantization can take advantage of many nonperturbative tools developed for field theories on non-dynamical backgrounds, whereas at the same time it is background independent in the sense that no special $\overline{g}_{\mu \nu}$ plays any distinguished role. During the quantization of the $h_{\mu \nu}$-field the background metric is kept fixed but is never specified explicitly.

In this construction the RG scale $k$ is ``proper'' with respect to the background metric. Technically one organizes the path integral over $h_{\mu \nu}$ according to eigenmodes of the covariant Laplacian $D^2 (\overline{g}_{\mu \nu})$ built from $\overline{g}_{\mu \nu}$ and cuts off the integration at the infrared (IR) scale $k^2$. This is done by adding a mode suppression term $\Delta_k S$ to the bare action. Hence $k$ is a $\overline{g}_{\mu \nu}$-proper momentum related to the scale set by the ``last mode integrated out'' and can be given an approximate physical meaning therefore. (See \cite{jan1,jan2} for a detailed discussion of this point.) This property of the gravitational average action is the central prerequisite for the effective field theory interpretation and for the possibility of performing ``RG improvements'' on the basis of $\Gamma_k$ \cite{bh,erick1,cosmo1,cosmo2,entropy,esposito,h1,h2,h3,girelli,mof}.

The price one has to pay for this advantage is that the average action is now a functional of two metrics:
$\Gamma_k [g_{\mu \nu}, \overline{g}_{\mu \nu}] \equiv
\Gamma_k [\,\overline{h}_{\mu \nu};\overline{g}_{\mu \nu}]$.
Here $g_{\mu \nu} \equiv \overline{g}_{\mu \nu} 
+ \overline{h}_{\mu \nu}$ is the expectation value of the microscopic metric. Only after having solved for the (now more complicated) RG flow of 
$\Gamma_k [g_{\mu \nu}, \overline{g}_{\mu \nu}]$ one can impose
$\overline{h}_{\mu \nu} =0$ and define the reduced functional
$\overline{\Gamma}_k [g_{\mu \nu}] \equiv
\Gamma_k [g_{\mu \nu}, g_{\mu \nu}]$
which generates the same on-shell matrix elements as the original one \cite{back}.

It should be stressed that the average action $\Gamma_k [\,\cdot\, , \, \cdot \,]$ and its RG flow are \textit{background independent} objects, in the sense of the word as it is used in loop quantum gravity \cite{A,R,T}, for instance. Both metrics, $g_{\mu \nu}$ and $\overline{g}_{\mu \nu}$, are just freely variable arguments and no metric plays any distinguished role. Furthermore, the mode cutoff is defined in terms of $D^2 (\,\overline{g}_{\mu \nu})$ which involves the variable metric $\overline{g}_{\mu \nu}$ and not any rigid one. This is in sharp contrast to matter field theories on a non-dynamical spacetime with a metric $g^{\text{non-dyn}}_{\mu \nu}$. There $\Delta_k S$ is constructed from $D^2 (g^{\text{non-dyn}}_{\mu \nu})$ which does indeed single out a specific metric. The resulting flow is not background independent in the above sense.

Besides fixing the physical scale of $k$, the use of the background field technique has a second, conceptually completely independent advantage: If one employs a gauge fixing term which is invariant under the background gauge transformations the resulting average action is a \textit{diffeomorphism invariant} functional of its arguments.

In the construction of the gravitational average action in \cite{mr} these two issues are intertwined and because of the complexity of realistic RG flows it is not easy to see how precisely the $\overline{g}_{\mu \nu}$-dependence of the IR cutoff
$\Delta_k S [h_{\mu \nu}; \overline{g}_{\mu \nu}]$ influences the flow. One of the purposes of the present paper is to study this influence in a setting as ``clean'' as possible, namely in an approximation to the full gravitational RG flow where gauge issues play no role and the impact of this $\overline{g}$-dependence of the cutoff can be studied in isolation. The implications of the $\overline{g}$-dependence are at the very heart of quantum gravity. It arises only because the metric has the crucial property, not shared by any other field, of defining the proper size of all dimensionful quantities, including that of $k$.

Within a different theory of gravity, and in a different formal setting, Floreanini and Percacci \cite{floper} have made similar observations. They studied a perturbatively renormalizable gauge theory of vielbein and spin connection fields. While asymptotic safety is not an issue there, they demonstrated that the quantization of the model results in a ``bimetric theory'', and depending on which metric is used in the ultraviolet (UV) regulator different effective potentials are obtained for the conformal factor.

The system we are going to study in the following obtains by approximating the gravitational RG flow in two ways: First, we restrict the theory space to that of the familiar Einstein--Hilbert truncation whose RG flow is known in full generality \cite{mr,frank1}. Second, we quantize only the conformal factor of the metric but not the other degrees of freedom it carries. This ``conformally reduced Einstein--Hilbert'' (or ``CREH'') truncation leads to a modified RG flow on the same theory space as the full Einstein--Hilbert truncation, and it will be very instructive to compare the two.

All metrics appearing in the CREH framework, the integration variable in the path integral, $\gamma_{\mu \nu}$, as well as $\overline{g}_{\mu \nu}$ and $g_{\mu \nu}$, are of the type ``conformal factor times $\widehat{g}_{\mu \nu}$'' where $\widehat{g}_{\mu \nu}$ is a reference metric which is never changed; for example, $\widehat{g}_{\mu \nu} = \delta_{\mu \nu}$. In this way, $\gamma_{\mu \nu}$, $\overline{g}_{\mu \nu}$,
and $g_{\mu \nu}$ get represented by a single ``scalar'' function, their respective conformal factor. The background metric, for instance, is written as $\overline{g}_{\mu \nu} = 
\overline{\phi}\,^2 (x) \, \widehat{g}_{\mu \nu}$. If one inserts the metric $\phi^2 \, \widehat{g}_{\mu \nu}$ into the Einstein--Hilbert action one obtains a $\phi^4$-type action for the field $\phi$, with a $\phi^4$-coupling proportional to the cosmological constant. We shall analyze this scalar--looking theory by means of an effective average action. We use a background approach which is analogous to the one used in the full gravitational FRGE. In particular the conformal factor of $\overline{g}_{\mu \nu}$ sets the physical scale of $k$. So, conceptually, this simplified setting is exactly the same as in the full gravitational flow equation, the only difference is that we allow only the quantum fluctuations of the conformal factor to contribute to the RG running of the couplings, i.\,e.\ the Newton and the cosmological constant, respectively.

The standard quantization of $\phi^4$-theory by means of an FRGE for the average action is fairly well understood \cite{avactrev}. It amounts to using a $\overline{g}_{\mu \nu}$-independent cutoff. Here $\Delta_k S$ is built from $\widehat{g}_{\mu \nu}$ which is usually taken to be the metric of flat Euclidean space. It is this metric $\widehat{g}_{\mu \nu}$ which defines the meaning of $k$. This scheme is the natural one when $\phi$ is a conventional scalar matter field. By now a lot is known about the resulting RG flow \cite{avactrev}. In particular, above all mass thresholds one recovers the $\ln (k)$-running of the $\phi^4$-coupling which is familiar from perturbation theory.

If $\phi$ is the conformal factor of the metric the situation is different. Now it is natural to define $\Delta_k S$ and hence $k$ in terms of the \textit{adjustable} background metric $\overline{g}_{\mu \nu} = 
\overline{\phi}\,^2 (x) \, \widehat{g}_{\mu \nu}$; its conformal factor $\overline{\phi}$ is determined dynamically by the condition that the fluctuations about $\overline{\phi}$ have vanishing expectation value. We shall see that the resulting RG flow is quite different from the standard scalar one. Typically one finds that the RG running is much faster in the gravitational case. For instance, there is a regime where the slow $\ln (k)$-running of the standard scalar is replaced by a much stronger $k^4$-running of the $\phi^4$-coupling. In this regime the $\phi^4$-coupling is proportional to the cosmological constant, $\Lambda_k$. Hence, in this particular regime, $\Lambda_k \propto k^4$.
This quartic cutoff dependence is something very well known, of course. It is precisely what one finds by summing zero-point energies, or rediscovers as quartic divergences in ordinary Feynman diagram calculations. Moreover it agrees with the result from the full Einstein--Hilbert truncation.

To summarize this important point: The (expected) behavior $\Lambda_k \propto k^4$ obtains only if we appreciate the very special role of gravity, namely that it determines all proper scales, including that of $k$. We find $\Lambda_k \propto k^4$ only if we define the cutoff with $\overline{g}_{\mu \nu} = 
\overline{\phi}\,^2 (x) \, \widehat{g}_{\mu \nu}$, while we obtain the much weaker $k$-dependence $\Lambda_k \propto \ln (k)$ if we treat $\phi$ as an ordinary scalar.

Earlier on Polyakov \cite{polyakov} and Jackiw et al.\ \cite{jackiw} have pointed out that in the CREH approximation the gravitational action is of the $\phi^4$-type and argued on the basis of standard scalar field theory that the cosmological constant should have a logarithmic scale dependence therefore. The results of the present paper indicate that if one wants to attach a physical meaning to $k$ by measuring it in units of $\phi$ itself the running of $\Lambda_k$ is much faster in fact.

Perhaps the most unexpected and striking feature of the CREH flow is that it admits a non-Gaussian RG fixed point (NGFP) with exactly the same qualitative properties as the one in the full Einstein--Hilbert truncation. The comparatively simple dynamics of a $\phi^4$-theory is enough to achieve asymptotic safety \textit{provided one takes into account that the field itself defines the proper coarse graining scale}.

At the NGFP the cosmological constant is positive and this translates to a negative $\phi^4$-coupling. Long ago Symanzik \cite{syman,hist} showed that the scalar $\phi^4$-theory with a negative coupling constant is asymptotically free; its coupling strength vanishes logarithmically at high momenta. Using the cutoff appropriate for the gravitational field the asymptotically free RG flow becomes an asymptotically safe one, a NGFP develops.

The investigations using the gravitational average action which have been performed during the past few years \cite{mr,percadou,oliver1,frank1,oliver2,oliver3,oliver4,souma,frank2,prop,oliverbook,perper1,codello,litimgrav} indicate that full Quantum Einstein Gravity (QEG) is indeed likely to possess a NGFP which makes the theory asymptotically safe. Increasingly complicated truncations of theory space were analyzed whereby all modes of the metric were retained. The results of the present paper indicate that the NGFP that was found in these analyses is perhaps easier to understand than it was thought up to now. It seems that, to some extent, it owes its existence to an essentially ``kinematical'' phenomenon which is related to the requirement of ``background independence'' and the fact that the dynamical field itself, the metric, determines the proper value of the coarse graining scale. The complicated selfinteractions of the helicity-2 modes, on the other hand, can be omitted without destroying the NGFP. While also characteristic of gravity, they seem not to be essential for asymptotic safety.

In this paper we shall see that all qualitative features of the RG flows derived from the CREH- and the full Einstein--Hilbert truncation, respectively, are almost identical. This gives rise to the hope that also in more general truncations the conformal factor plays a representative role so that, with some care, we might be able to learn something about the full gravitational dynamics from the conformal reduction. Technically the ``scalar'' theory is comparatively simple so that we should be able to perform computations in regions of theory space which are unaccessible otherwise. We start this program in the companion paper \cite{creh2} where we analyze the local potential approximation (LPA) for $\phi$; it involves a running potential $U_k (\phi)$ with an arbitrary field dependence.

The remaining sections of this paper are organized as follows. As a preparation we discuss in Section \ref{s2} the conformally reduced Einstein--Hilbert action. Then, in Section \ref{s3}, we derive an exact flow equation for conformally reduced gravity. In Section \ref{s4} we specialize it for the CREH truncation and explain in particular the conceptual differences of the theory presented here and standard scalar matter field theories. In Section \ref{s5} we analyze the RG equations obtained from the CREH truncation and show that they predict a NGFP. The conclusions are contained in Section \ref{s6}.
%
%
%
%
%
%
\section{The Conformally Reduced Einstein--Hilbert Action}\label{s2}
In $d$ spacetime dimensions, the Euclidean Einstein--Hilbert action reads
\begin{align} \label{2.1}
S_{\text{EH}} [g_{\mu \nu}]
& =
- \frac{1}{16 \pi \, G} \,
\int \!\!\mathrm{d}^d x~\sqrt{g\,} \,
\bigl( R (g) - 2 \, \Lambda \bigr).
\end{align}
Henceforth we shall assume that the argument $g_{\mu \nu}$ is a conformal factor times a fixed, non-dynamical reference metric $\widehat{g}_{\mu \nu}$. We would like to parameterize this conformal factor in terms of a ``scalar'' function $\phi (x)$ in such a way that the kinetic term for $\phi$ becomes standard,
$\propto (\partial_\mu \phi)^2$. This is indeed possible for any dimensionality. Introducing $\phi$ according to \cite{jackiw},
\begin{align} \label{2.2}
g_{\mu \nu} 
& =
\phi^{2 \nu (d)} \, \,\widehat{g}_{\mu \nu},
\end{align}
with the exponent
\begin{align} \label{2.3}
\nu (d) & \equiv \frac{2}{d-2}
\end{align}
standard formulas for Weyl rescalings yield the following result for $S_{\text{EH}}$ evaluated on metrics of the form \eqref{2.2}:
\begin{align} \label{2.4}
S_{\text{EH}} [\phi]
& =
- \frac{1}{8 \pi \, \xi (d) \, G} \,
\int \!\! \mathrm{d}^d x~ \sqrt{\widehat{g} \,} \,
\left(
\tfrac{1}{2} \, \widehat{g}\,^{\mu \nu} \, \partial_\mu \phi \, 
\partial_\nu \phi
+ \tfrac{1}{2} \, \xi (d) \, \widehat{R} \, \phi^2
- \xi (d) \, \Lambda \, \phi^{2d/(d-2)} 
\right).
\end{align}
Here $\widehat{R}$ is the curvature scalar of the reference metric $\widehat{g}_{\mu \nu}$, and
\begin{align} \label{2.5}
\xi (d) & \equiv \frac{d-2}{4 \, \left( d-1 \right)}.
\end{align}
In $4$ dimensions we have $\nu = 1$ and $\xi = 1/6$ so that the choice
\begin{align} \label{2.6}
g_{\mu \nu} & = \phi^2 \, \,\widehat{g}_{\mu \nu}
\end{align}
converts the Einstein--Hilbert action to a kind of ``$\phi^4$-theory'':
\begin{align} \label{2.7}
S_{\text{EH}} [\phi]
& =
- \frac{3}{4 \pi \, G} \,
\int \!\! \mathrm{d}^4 x~ \sqrt{\widehat{g} \,} \,
\left(
\tfrac{1}{2} \, \widehat{g}\,^{\mu \nu} \, \partial_\mu \phi \, 
\partial_\nu \phi
+ \tfrac{1}{12} \, \widehat{R} \, \phi^2
- \tfrac{1}{6} \, \Lambda \, \phi^4 
\right).
\end{align}
We shall refer to the action \eqref{2.4} and its special case \eqref{2.7} as the conformally reduced Einstein--Hilbert or ``CREH'' action.

Several comments are in order here. 

\noindent(a)~Up to now $\widehat{g}_{\mu \nu}$ is an arbitrary metric, defined on the same smooth manifold as $g_{\mu \nu}$. Later on we shall fix the topology of this manifold to be that of flat space $\mathrm{R}^d$ or of the sphere $\mathrm{S}^d$.

\noindent(b)~For $d>2$, the case we shall always assume in the following, the kinetic term in $S_{\text{EH}} [\phi]$ of eq.\ \eqref{2.4} is always negative definite due to the ``wrong sign'' of its prefactor. As a result, the action is unbounded below: for a $\phi (x)$ which varies sufficiently rapidly $S_{\text{EH}} [\phi]$ can become arbitrarily negative. This is the notorious conformal factor instability.

\noindent(c)~To make the analogy with a scalar action perfect one could remove the prefactor $1 / ( 8 \pi \xi (d) G )$ in eq.\ \eqref{2.4} by a rescaling of $\phi$. (But not the minus sign!) We shall not perform this rescaling here because later on Newton's constant will depend on the RG scale $k$, and $k$-dependent field rescalings are not allowed in the FRGE formalism we shall use.

\noindent(d)~The potential terms in $S_{\text{EH}} [\phi]$ are a mass--type term proportional to the scalar curvature of $\widehat{g}_{\mu \nu}$ and an interaction term $\propto \phi^{2d/(d-2)}$. The selfinteraction of the $\phi$-field is entirely due to the cosmological constant; it vanishes for $\Lambda =0$. The exponent $2d/(d-2)$ equals $4$ in $d=4$, is smaller than $4$ in higher dimensions (approaching $2$ for $d \to \infty$) and larger than $4$ in lower dimensions (diverging to $+ \infty$ for $d \searrow 2$).

Leaving aside issues related to the functional measure, quantizing gravity in the CREH approximation based upon the bare action $S_{\text{EH}} [\phi]$ is similar to quantizing a scalar theory with an action of the general type
\begin{align} \label{2.8}
S [\phi]
& =
c \, \int \!\! \mathrm{d}^4 x ~
\Big\{ - \tfrac{1}{2} \left( \partial \phi \right)^2
+ U (\phi) \Big\}
\end{align}
where $c$ is a positive constant.
For the sake of the argument let us assume that 
$\widehat{g}_{\mu \nu} = \delta_{\mu \nu}$ is the flat metric on $\mathrm{R}^4$. Then $S_{\text{EH}}$ of \eqref{2.7} is indeed of the form \eqref{2.8} with the potential 
$U (\phi) = \tfrac{1}{6} \, \Lambda \, \phi^4$ and
$c = 3/ (4 \pi G) >0$. Let us assume that the cosmological constant is positive, the case which will be relevant later on. For $\Lambda >0$ the potential term in the action \eqref{2.8} is positive definite, while the kinetic piece is negative definite. We would like to explore the quantum theory based upon the functional integral
\begin{align} \label{2.9}
I & \equiv \int \!\! \mathcal{D} \phi~
\ee^{i \widetilde{S} [\phi]}
\end{align}
where $\widetilde{S}$ is the Wick rotated version of $S$, with
$(\partial \phi)^2 \equiv \eta^{\mu \nu} \, \partial_\mu \phi \,
\partial_\nu \phi$. One would expect that in this theory the wrong sign of the kinetic term drives the condensation of spatially inhomogeneous ($x$-dependent) modes, i.\,e.\ the formation of a ``kinetic condensate'' similar to the one discussed in \cite{kincond}. The amplitude of the inhomogeneous modes cannot grow unboundedly since this would cost potential energy.

Next let us look at the closely related theory with the ``inverted'' action $S_{\text{inv}} [\phi] \equiv - S [\phi]$. Thus
\begin{align} \label{2.10}
S_{\text{inv}} [\phi]
& =
c \, \int \!\! \mathrm{d}^4 x ~
\Big\{ + \tfrac{1}{2} \left( \partial \phi \right)^2
+ V (\phi) \Big\}
\end{align}
with the negative potential
\begin{align} \label{2.11}
V (\phi) & \equiv - U (\phi) \leq 0.
\end{align}
In pulling out a global minus sign from $S$ the instability inherent in the theory has been shifted from the kinetic to the potential term. According to $S_{\text{inv}}$, the kinetic energy assumes its minimum for homogeneous configurations $\phi = const$, but the inverted potential $V (\phi) = - \tfrac{1}{6} \, \Lambda \, \phi^4$ becomes arbitrarily negative for large $\phi$.

Even though $S$ and $S_{\text{inv}}$ appear to be plagued by instabilities of a very different nature, they nevertheless describe the same physics (up to a time reflection). The path integrals involving $S$ and $S_{\text{inv}}$ are related by a simple complex conjugation:
\begin{align} \label{2.12}
I_{\text{inv}}
& \equiv 
\int \!\! \mathcal{D} \phi~
\ee^{- i \widetilde{S} [\phi]}
= I^\ast.
\end{align}
We shall refer to the formulation in terms of $S$ and $S_{\text{inv}}$ as the \textit{original picture} and the \textit{inverted picture}, respectively.

So we see that for \textit{pure} gravity in the CREH approximation the ``wrong'' sign of the kinetic term can be traded for an upside down potential. The FRGE formalism we are going to develop will effectively correspond to the inverted picture. As we shall see it is indeed the $\Lambda>0$ case that will be relevant to asymptotic safety. Hence the conformal factor dynamics is described by an action with positive kinetic but negative potential term.

Interestingly enough, this kind of $\phi^4$-theory with a negative coupling constant was discussed by Symanzik \cite{syman} long ago. He showed that the coupling strength vanishes at short distances, thus providing the first example of an asymptotically free quantum field theory \cite{hist}.
%
%
%
\section{Effective Average Action for the Conformal Factor}\label{s3}
\subsection{The Background Field Method}\label{s3.1}
We emphasized already repeatedly that the most important difference between the conformal factor and an ordinary scalar is that $\phi$ determines the magnitude of all physical scales; in particular it determines the proper scale that is to be ascribed to a given numerical value of the IR cutoff $k$ appearing in the FRGE context. For this reason the quantization of $\phi$ by means of an FRGE differs from the standard one. In fact, even though gauge issues do not play any role here, the background field method has to be employed. This approach will allow us to give a meaning to statements like ``$\Gamma_k$ describes the dynamics of fields averaged over spacetime volumes of extension $\sim k^{-1}$''
in presence of a quantized metric where a priori it is unclear in which metric the extension of those spacetime volumes is measured.

Before we can set up the RG formalism we must explain the background--reformulation of the path integral underlying the quantum field theory of the conformal factor. We start from a formal path integral\footnote{Since this is customary in the literature we shall use an Euclidean notation in the general discussions. At the formal level it its trivial to obtain the corresponding Lorentzian formulas by replacing $-S \to iS$, etc.; for the time being the positivity properties of $S$ play no role.}
\begin{align} \label{3.1}
\int \!\! \mathcal{D} \chi ~\ee^{-S [\chi]}
\end{align} 
where $S$ is an arbitrary bare action (perhaps related, but not necessarily identical to $S_{\text{EH}}$) and $\chi (x)$ denotes the microscopic (``quantum'') conformal factor field. (The notation $\phi (x)$ will be reserved for its expectation value.) We think of \eqref{3.1} as descending from a path integral over quantum metrics $\gamma_{\mu \nu} (x)$,
\begin{align} \label{3.2}
\int \!\! \mathcal{D} \gamma_{\mu \nu}~
\ee^{-S_{\text{grav}} [\gamma_{\mu \nu}]},
\end{align}
by a restriction to metrics of the form
\begin{align} \label{3.3}
\gamma_{\mu \nu} & = \chi^{2 \nu} \, \,\widehat{g}_{\mu \nu}.
\end{align}
The integrals \eqref{3.1} and \eqref{3.2} refer to a spacetime manifold of a given topology and $\widehat{g}_{\mu \nu}$ is a reference metric consistent with this topology. The action $S [\chi]$ depends parametrically on $\widehat{g}_{\mu \nu}$ but we shall not indicate this dependence notationally. The non-dynamical, classical metric $\widehat{g}_{\mu \nu}$ is considered fixed once and for all; it has no analog in the full theory and is not to be confused with the background metric and the corresponding conformal factor which we introduce next.

We decompose the variable of integration, $\chi$, as the sum of a classical, fixed background field $\chib$ and a fluctuation $f$:
\begin{align} \label{3.4}
\chi (x) & = \chib (x) + f (x).
\end{align}
Even though we frequently use the term ``fluctuation'', $f (x)$ is not assumed small, and no expansion in powers of $f (x)$ is performed here.
We assume that the measure $\mathcal{D} \chi$ is translational invariant so that \eqref{3.1} can be replaced by $\int \! \mathcal{D} f \, \exp (-S[\chib +f])$. Actually it is sufficient to assume that the original $\mathcal{D} \chi$ equals a translational invariant measure up to a Jacobian since we may include the logarithm of this Jacobian in $S$.

At this point it is natural to introduce a background--type generating functional by coupling an external source $J (x)$ to the fluctuation only:
\begin{align} \label{3.5}
\exp \bigl( W [J; \chib] \bigr)
& =
\int \!\! \mathcal{D} f ~
\exp \left( -S [\chib +f]
+ \int \!\! \mathrm{d}^d x~\sqrt{\widehat{g}\,} \,
J (x) \, f (x) \right).
\end{align}
Repeated differentiation of $W$ with respect to the source yields the connected $n$-point functions of $f$ in presence of $J$. In particular the normalized expectation value of the fluctuation is
\begin{align} \label{3.6}
\overline{f} (x) & \equiv \langle f (x) \rangle
= \frac{1}{\sqrt{\widehat{g} (x)\,}\,} \,
\frac{\delta W[J; \chib]}{\delta J (x)}.
\end{align}
The field thus obtained is functionally dependent on both $J$ and $\chib$, i.\,e.\ $\overline{f} = \overline{f} [J; \chib]$. We assume that this relationship can be solved for the source, $J = J[\,\overline{f}; \chib]$, and introduce the Legendre transform of $W$:
\begin{align} \label{3.7}
\Gamma [\,\overline{f}; \chib]
& =
\int \!\! \mathrm{d}^d x~\sqrt{\widehat{g}\,} \,
J [\,\overline{f}; \chib] (x) \, \overline{f} (x)
- W \bigl[ J [\,\overline{f}; \chib]; \chib \bigr].
\end{align}
This definition implies the effective field equation
\begin{align} \label{3.8}
\frac{\delta \Gamma [\,\overline{f}; \chib]}{\delta \overline{f} (x)}
& = J (x).
\end{align}
More generally, repeated differentiation of $\Gamma$ with respect to $\overline{f} (x)$ yields the 1PI $n$-point correlators of $\overline{f}$ in presence of $J$. The source can be ``switched off'' by equating $\overline{f}$ after the differentiations to the function $\overline{f}_0 [\chib] (x) \equiv \overline{f} \,[J=0; \chib] (x)$. Note that $\overline{f}_0$ has no reason to vanish in general, and that the resulting $n$-point functions still depend on $\chib$. The expectation value of the complete conformal factor reads
\begin{align} \label{3.9}
\phi & \equiv \big\langle \left( \chib + f \right) \big\rangle
= \chib + \overline{f},
\end{align}
and sometimes it will be convenient to regard $\Gamma$ a 
functional of $\phi$ and $\chib$ rather than $\overline{f}$ and $\chib$:
\begin{align} \label{3.10}
\Gamma [\phi, \chib] & \equiv 
\Gamma [\,\overline{f} = \phi - \chib;\,\chib].
\end{align}
For the restriction of this function to equal arguments $\phi = \chib$ which amounts to a vanishing fluctuation expectation value we write
\begin{align} \label{3.11}
\overline{\Gamma} \,[\phi] & \equiv \Gamma [\phi, \phi]
= \Gamma [\,\overline{f} =0;\,\chib=\phi].
\end{align}

It is instructive to compare the above generating functionals in the background approach with those in the standard (``st''), i.\,e.\ non-background formalism. There one would define $W_{\text{st}} [J]$ by
\begin{align} \label{3.12}
\exp \bigl( W_{\text{st}} [J] \bigr)
& =
\int \!\! \mathcal{D} \chi \,
\exp \left( - S [\chi] 
+ \int \!\! \mathrm{d}^d x~ \sqrt{\widehat{g}\,} \,
J (x) \, \chi (x) \right)
\end{align}
and the standard effective action $\Gamma_{\text{st}} [\phi]$ would obtain as the Legendre transform of $W_{\text{st}} [J]$. Exploiting the translational invariance of $\mathcal{D} \chi$ it is easy to see that the two sets of functionals are related in a rather trivial way:
\begin{gather}
\label{3.13}
W [J; \chib] 
= 
W_{\text{st}} [J] - \int \!\! \mathrm{d}^d x~\sqrt{\widehat{g}\,} \,
J (x) \, \chib (x)
\\
\label{3.14}
\Gamma [\,\overline{f}; \chib]
=
\Gamma_{\text{st}} [\chib+\overline{f}\,]
\quad \Longleftrightarrow \quad
\Gamma [\phi, \chib] = \Gamma_{\text{st}} [\phi]
\\
\label{3.15}
\overline{\Gamma}\, [\phi] 
= \Gamma_{\text{st}} [\phi].
\end{gather}
The key property of the background formalism is that the standard $n$-point functions
\begin{align} \label{3.16}
\frac{\delta^n \Gamma_{\text{st}} [\phi]}
{\delta \phi (x_1) \dotsm \delta \phi (x_n)}
\end{align}
can alternatively be computed by differentiating the functional $\Gamma [\,\overline{f}=0; \,\chib] = \overline{\Gamma} \,[\chib]$ with respect to the background $\chib$:
\begin{align} \label{3.17}
\frac{\delta^n \Gamma [\,\overline{f}=0; \,\chib]}
{\delta \chib (x_1) \dotsm \delta \chib (x_n)}
\Bigg \rvert_{\chi_{\text{B}}=\phi}
& \equiv
\frac{\delta^n \overline{\Gamma}_{\text{st}} [\phi]}
{\delta \phi (x_1) \dotsm \delta \phi (x_n)}.
\end{align}
In the case at hand the equality of \eqref{3.16} and \eqref{3.17} is trivial since $\Gamma [\,\overline{f}=0;\,\chib]$ and $\Gamma_{\text{st}} [\chib]$ are exactly equal here.

The situation is less trivial when one applies this formalism to gauge theories, employing a gauge fixing term invariant under background gauge transformations. Then the analogs of the $n$-point functions \eqref{3.16} and \eqref{3.17} are not exactly equal, but they are equal ``on-shell''. As a result, both sets of correlators give rise to the same set of physical $S$-matrix elements \cite{back}. 
The important conclusion is that even then the functional $\overline{\Gamma}$ which obtains by requiring that the fluctuation has no expectation value (\,$\overline{f} =0$) and depends only on one field ($\chib \equiv \phi$) contains all of the physical, gauge--invariant information.

Before continuing let us summarize the status of the various metrics, all conformal to one another, that enter the construction. First, there is the \textit{reference metric} $\widehat{g}_{\mu \nu}$, a classical field which is fixed once and for all and never gets varied. Second, there is the \textit{quantum metric}, the integration variable
\begin{align} \label{3.18}
\gamma_{\mu \nu} 
& = \chi^{2 \nu} \, \widehat{g}_{\mu \nu}
= \left( \chib + f \right)^{2 \nu} \, \,\widehat{g}_{\mu \nu}.
\end{align}
In the canonical approach this metric corresponds to an operator. Third, there is the \textit{background metric} defined by
\begin{align} \label{3.19}
\overline{g}_{\mu \nu} 
& \equiv \chib^{2 \nu} \, \,\widehat{g}_{\mu \nu}.
\end{align}
It is a classical field again which is considered variable, however. In particular it can be adjusted to achieve $\overline{f} =0$ if this is desired. Fourth, there is the \textit{expectation value of the quantum metric}
\begin{align} \label{3.20}
g_{\mu \nu} & \equiv \langle \gamma_{\mu \nu} \rangle 
\equiv
\left\langle \left( \chib + f \right)^{2 \nu} \right\rangle 
\, \widehat{g}_{\mu \nu}.
\end{align}
And finally, fifth, there is the \textit{metric with the conformal factor $\phi$}. As $\phi \equiv \chib + \overline{f} = \chib + \langle f \rangle$, it reads
\begin{align} \label{3.21}
\breve{g}_{\mu \nu}
& \equiv
\phi^{2 \nu} \, \,\widehat{g}_{\mu \nu}
\equiv
\bigl( \chib + \left\langle f \right\rangle \bigr)^{2 \nu} \, \,
\widehat{g}_{\mu \nu}.
\end{align}
In general $g_{\mu \nu}$ and $\breve{g}_{\mu \nu}$ are not exactly equal. However, they are approximately equal if the quantum fluctuations of $f$ are small. In $d=4$ where $\nu=1$, for instance, we have
\begin{align} \label{3.22}
\begin{split}
g_{\mu \nu}
& =
\overline{g}_{\mu \nu} 
+ \left[ 2 \, \chib \, \left\langle f \right\rangle
+ \left\langle f^2 \right\rangle
\right] \, \widehat{g}_{\mu \nu}
\\
\breve{g}_{\mu \nu}
& =
\overline{g}_{\mu \nu} 
+ \left[ 2 \, \chib \, \left\langle f \right\rangle
+ \left\langle f \right\rangle^2
\right] \, \widehat{g}_{\mu \nu}.
\end{split}
\end{align}
Hence the difference $g_{\mu \nu} - \breve{g}_{\mu \nu} =
\left[ \langle f^2 \rangle - \langle f \rangle^2 \right] \,
\widehat{g}_{\mu \nu}$ is proportional to the variance of $f$ so that $g_{\mu \nu}$ and $\breve{g}_{\mu \nu}$ are not very different if the fluctuations of $f$ are ``small''. However, in order to make this statement precise one first would have to give a meaning to the expectation value of the operator product $f^2$ with both operators at the same point, something we shall not attempt here. Notice also that $\breve{g}_{\mu \nu}$ reduces to $\overline{g}_{\mu \nu}$ if $\overline{f} =0$ while $g_{\mu \nu}$ does not: $g_{\mu \nu} = \overline{g}_{\mu \nu} + \langle f^2 \rangle \, \widehat{g}_{\mu\nu}$.

The metrics $g_{\mu \nu}$ and $\overline{g}_{\mu \nu}$ are analogous to the fields with the same names in the construction of the exact gravitational average action \cite{mr}. Certain differences arise, however, since there a linear background--quantum split is performed at the level of the full metric, while in the present approach the split is linear at the level of the conformal factor. In ref.\ \cite{mr} where the integral over all metrics $\gamma_{\mu \nu}$ is dealt with, one decomposes $\gamma_{\mu \nu} = \overline{g}_{\mu\nu} + h_{\mu \nu}$ and then integrates over the fluctuation $h_{\mu \nu}$. As a result, $g_{\mu \nu} = \langle \gamma_{\mu \nu} \rangle = \overline{g}_{\mu\nu} + \langle h_{\mu \nu} \rangle$ is linear in the expectation value of the fluctuation so that there is no difference between $g_{\mu \nu}$ and $\breve{g}_{\mu \nu}$. In the present setting, on the other hand, the metric $\gamma_{\mu \nu}$ is parameterized by the fluctuation in a \textit{nonlinear} way: $\gamma_{\mu \nu} = (\chib + f)^{2 \nu} \, \widehat{g}_{\mu \nu}$. This nonlinearity is the price we have to pay if we want the CREH action to look like that of a standard scalar $\phi^4$-theory. In the present paper we are mostly interested in understanding the differences between the dynamics of the conformal factor and that of a standard scalar, and we are therefore going to accept this nonlinearity. As a consequence, the calculations within the ``CREH truncation'' we are going to perform later on are not simply a subset of the calculations in the full Einstein--Hilbert truncation which includes the transverse etc.\ modes of the metric in addition to the conformal factor.
\subsection{The Average Action of the Conformal Factor}\label{s3.2}
From the technical point of view, the main problem consists in (approximately) computing the path integral \eqref{3.5}. Next we shall set up an RG formalism which translates this problem into the equivalent problem of solving a certain functional RG equation subject to a boundary condition involving $S$. Using a variant of the effective average action for scalars \cite{avact,livrev} we modify \eqref{3.5} by introducing a mode--cutoff term into the path integral defining $W$:
\begin{align} \label{3.23}
\begin{split}
& \exp \bigl( W_k [J; \chib] \bigr)
\\
& \phantom{{=}} =
\int \!\! \mathcal{D} f~
\exp \left( - S [\chib +f] - \Delta_k S [f; \chib]
+ \int \!\! \mathrm{d}^d x ~\sqrt{\widehat{g}\,} \,
J (x) \, f (x) \right).
\end{split}
\end{align}
The action $\Delta_k S [f; \chib]$ is to be constructed in such a way that the factor $\exp (- \Delta_k S)$ suppresses the long--wavelength modes of $f (x)$ with momenta $p \lesssim k$ while it does not affect the short--wavelength modes with $p \gtrsim k$. In order to arrive at an FRGE of the familiar second--order type we take $\Delta_k S$ to be quadratic in $f$:
\begin{align} \label{3.24}
\Delta_k S [f; \chib]
& = 
\tfrac{1}{2} \, \int \!\! \mathrm{d}^d x~\sqrt{\widehat{g}\,} \,
f (x) \, \mathcal{R}_k [\chib] \, f (x).
\end{align}
Here $\mathcal{R}_k$ is a pseudodifferential operator which may depend on the background field. Allowing for this $\chib$-dependence is crucial in order to implement ``background independence'' \cite{A,R,T} and to give a ``proper'' meaning to the coarse graining scale $k$ in a theory with a dynamical metric.

In flat space the parameter $k$, by elementary Fourier theory, has the interpretation of the inverse length scale over which the microscopic fields are averaged or ``coarse grained''. If we want to have a similar interpretation in quantum gravity we must decide with respect to which metric this length scale is measured. In the background field approach, there is a canonical candidate for a metric measuring the coarse graining scale, namely the background metric $\overline{g}_{\mu \nu} = \chib^{2 \nu} \, \widehat{g}_{\mu\nu}$. In fact, as we discussed in the Introduction, the RG flow becomes ``background independent'' (in the sense of \cite{A,R,T}) if $\Delta_k S$ is constructed from $\overline{g}_{\mu \nu}$, or $\chib$ here, rather than from a rigid metric. The key property of $\mathcal{R}_k [\chib]$ is to distinguish ``long--wavelength'' and ``short--\-wave\-length'' modes of $f (x)$ whereby the ``length'' is defined in terms of $\overline{g}_{\mu \nu}$, i.\,e.\ the background conformal factor $\chib$.

The advantage of using the background field method is that at an intermediate stage it decouples the field integrated over, the fluctuation $f$, from the field that fixes the physical value of $k$, namely $\chib$. At the very end, after the quantization has been performed and the RG trajectories are known, we may set $\overline{f} =0$ without loosing information. Then the scale dependent version of the single--argument functional defined above, $\overline{\Gamma}_k [\phi \equiv \chib]$, depends only on one conformal factor, corresponding to ``the'' metric $g_{\mu \nu}$, and its parameter $k$ is a momentum measured, indirectly, with respect to this metric.

A cutoff operator $\mathcal{R}_k$ with the desired properties can be constructed along the following lines. We think of the functional integral \eqref{3.23} over $f$ as being organized according to eigenfunctions of the Laplace--Beltrami operator constructed from $\overline{g}_{\mu \nu}$:
\begin{align} \label{3.25}
\overline{\Box} & \equiv \overline{g}\,^{-1/2} \,
\partial_\mu \, \overline{g}\,^{1/2} \, 
\overline{g}\,^{\mu \nu} \, \partial_\nu.
\end{align}
Expanding $f$ in terms of $(- \overline{\Box}\,)$-eigenfunctions, the task of $\mathcal{R}_k$ is to suppress those with eigenvalues smaller than $k^2$ by giving them a ``mass'' of the order $k$, while those with larger eigenvalues must remain ``massless'' \cite{avact,livrev}. In the simplest case when the $f$-modes have a kinetic operator proportional to $\overline{\Box}$ itself the rule is that the correct $\mathcal{R}_k$ when added to $\Gamma_k^{(2)}$ leads to the replacement
\begin{align} \label{3.26}
(- \overline{\Box}\,)
& \longrightarrow
(- \overline{\Box} \,) 
+ k^2 \, R^{(0)} \bigl( \tfrac{- \overline{\Box}\,}{k^2\,} \bigr).
\end{align}
Here $R^{(0)} (z)$ is an arbitrary ``shape function'' interpolating between $R^{(0)} (0) =1$ and $R^{(0)} (\infty) =0$, with a transition region centered around $z=1$. These conditions guarantee that the effective inverse propagator of the long-- and short--wavelength modes is $- \overline{\Box} + k^2$ and $- \overline{\Box}$, respectively, and that the long/short--transition is at the $- \overline{\Box}$-eigenvalue $k^2$, as it should be.

The coarse graining scale $\ell = \ell (k)$ corresponding to the cutoff value $k$ is found by investigating the properties of the $-\overline{\Box}$-eigenfunction with eigenvalue $k^2$, the so-called ``cutoff mode'' \cite{jan1,jan2}: one determines its typical scale of variation with respect to $x$ (a period, say) and converts this coordinate length to a physical, i.\,e.\ proper length using $\overline{g}_{\mu \nu}$. The result, $\ell (k)$, is an approximate measure for the extension of the spacetime volumes up to which the dynamics has been ``coarse grained''. If $\overline{g}_{\mu \nu}$ is close to a flat metric, $\ell (k)$ equals approximately $\pi / k$. (See \cite{jan1,jan2} for a detailed discussion.) It is in this sense that the background metric $\overline{g}_{\mu \nu}$, or rather its conformal factor $\chib$, determines the physical (proper) scale of $k$.

Defining the scale $k$ as a cutoff in the spectrum of the covariant Laplacian built from $\overline{g}_{\mu \nu}$ is in accord with the construction of the exact gravitational average action in \cite{mr}; there, too, it is the background metric which sets the scale of $k$.

While this choice appears very natural, and in fact is the only meaningful one in the gravitational context, every standard quantization and RG scheme which treats $\phi$ as an ordinary scalar, at least implicitly, uses a differently defined cutoff, namely one based upon $\widehat{\Box}$. Here $\widehat{\Box}$ denotes the Laplace--Beltrami operator pertaining to the reference metric, 
$\widehat{\Box} =\widehat{g}\,^{-1/2} \,
\partial_\mu \, \widehat{g}\,^{1/2} \, 
\widehat{g}\,^{\mu \nu} \, \partial_\nu$, and $\mathcal{R}_k$ is designed to implement the replacement
\begin{align} \label{3.27}
(- \widehat{\Box})
& \longrightarrow
(- \widehat{\Box}) 
+ k^2 \, R^{(0)} \bigl( \tfrac{- \widehat{\Box}}{k^2\,} \bigr).
\end{align}
In this case the proper scale of $k$ is determined by the metric $\widehat{g}_{\mu \nu}$ which, however, at no stage of the construction acquires any physical meaning. As we emphasized, $\widehat{g}_{\mu \nu}$ is never varied. It ``knows'' nothing about the true (``on-shell'') metric of spacetime, namely the particular background metric which adjusts itself dynamically upon setting $\langle f \rangle =0$. The scheme \eqref{3.27} is the correct choice if one considers $\chi$ a standard scalar field on a non-dynamical spacetime with metric $\widehat{g}_{\mu \nu}$, on flat space ($\widehat{g}_{\mu \nu}=\delta_{\mu \nu}$), for instance. The average action formalism based upon \eqref{3.27} reproduces all the familiar results of perturbation theory, the $\ln (k)$-running of the quartic coupling in $\phi^4$-theory, for instance.

Since $\widehat{g}_{\mu \nu}$ is a rigid metric, the flow resulting from the substitution \eqref{3.27} is not ``background independent'' in the sense of \cite{A,R,T}, while \eqref{3.26} does indeed give rise to a ``background independent'' RG flow.

As we shall see, the flow based upon the $\overline{\Box}$-scheme \eqref{3.26} is \textit{extremely} different from the one for standard scalars. The reason is, of course, that via the $\chib$-dependence of $\overline{\Box}$ the gravitational field itself sets the scale of $k$. The difference between \eqref{3.26} and \eqref{3.27} becomes manifest when we recall that the Laplacians of 
$\widehat{g}_{\mu \nu}$ and 
$\overline{g}_{\mu \nu} = \chib^{2 \nu} \, \,\widehat{g}_{\mu \nu}$ are related by
\begin{align} \label{3.28}
\overline{\Box} 
& =
\chib^{-2 \nu} \, \widehat{\Box}
+ \mathcal{O} (\partial \chib).
\end{align}
The factor $\chib^{-2 \nu}$ leads to dramatic modifications of the RG flow whereas the $\mathcal{O} (\partial \chib)$-terms are less important; within the Einstein--Hilbert truncation they play no role.

After having discussed $\mathcal{R}_k$, the remaining steps of the construction follow the familiar rules \cite{avact,avactrev,livrev}. One defines the $k$-dependent field expectation value
\begin{align} \label{3.29}
\overline{f} (x) & \equiv \big\langle f(x) \big\rangle_k
=
\frac{1}{\sqrt{\widehat{g} (x)\,}\,} \,
\frac{\delta W_k [J; \chib]}{\delta J (x)},
\end{align}
solves for the source, $J (x) = J_k [\,\overline{f}; \chib] (x)$, and finally defines the effective average action $\Gamma_k$ as the Legendre transform of $W_k$ with $\Delta_k S [\,\overline{f}; \chib]$ subtracted:
\begin{align} \label{3.30}
\begin{split}
\Gamma_k [\,\overline{f}; \chib]
& =
\int \!\! \mathrm{d}^d x ~\sqrt{\widehat{g}\,} \,\,
\overline{f} (x) \, J_k [\,\overline{f}; \chib] (x)
- W_k \bigl[ J_k [\,\overline{f}; \chib]; \chib \bigr]
\\
& \phantom{{==}}
- \tfrac{1}{2} \, \int \!\! \mathrm{d}^d x ~\sqrt{\widehat{g}\,} \,\,
\overline{f} \, \mathcal{R}_k [\chib] \, \overline{f}.
\end{split}
\end{align}
In analogy with \eqref{3.10} and \eqref{3.11} we also introduce
\begin{gather}
\label{3.31}
\Gamma_k [\phi, \chib]
\equiv
\Gamma_k [\,\overline{f}=\phi-\chib;\,\chib]
\\
\label{3.32}
\overline{\Gamma}_k [\phi]\
\equiv \Gamma_k [\phi, \phi]
= 
\Gamma_k [\,\overline{f}=0;\,\chib=\phi].
\end{gather}

The main properties of $\Gamma_k$ are easily established along the same lines as in standard scalar theories \cite{avact,avactrev,livrev}. In particular, differentiating \eqref{3.23} with respect to $k$ leads to the following FRGE which governs the scale dependence of $\Gamma_k$:
\begin{align} \label{3.33}
k \partial_k \, \Gamma_k [\,\overline{f};\chib]
& = \tfrac{1}{2} \, \tr
\left[ \left( \Gamma_k^{(2)} [\,\overline{f};\chib]
+ \mathcal{R}_k [\chib] \right)^{-1} \,
k \partial_k \, \mathcal{R}_k [\chib] \right].
\end{align}
Here $\Gamma_k^{(2)}$ is the matrix of second functional derivatives of $\Gamma_k [\,\overline{f};\chib]$ with respect to $\overline{f}$ at fixed $\chib$. In bra--ket notation,
\begin{align} \label{3.34}
\langle x \lvert \Gamma_k^{(2)} \rvert y \rangle
& =
\frac{1}{\sqrt{\widehat{g} (x)\,} \, \sqrt{\widehat{g} (y)\,}\,}
\, \frac{\delta^2 \Gamma_k [\,\overline{f};\chib]}
{\delta \overline{f} (x) \, \delta \overline{f} (y)}.
\end{align}
Note that the metric appearing in formulas such as \eqref{3.29} or \eqref{3.32} is $\widehat{g}_{\mu \nu}$ (and not $\overline{g}_{\mu \nu}$). Correspondingly $\tr (\cdots) \equiv \int \! \mathrm{d}^d x~
\sqrt{\widehat{g}\,} \, \langle x \lvert (\cdots) \rvert x \rangle$. Notice also that, since the $\overline{f}$-derivatives are to be performed at fixed $\chib$, the FRGE \eqref{3.34} cannot be formulated in terms of the single--argument functional $\overline{\Gamma}_k$ alone. Hence the relevant theory space consists of functionals depending on two fields, $\overline{f}$ and $\chib$, or alternatively $\phi$ and $\chib$.

By construction $\mathcal{R}_k$ vanishes for $k \to 0$. As a consequence, $\Gamma_k$ reduces to the ordinary effective action in this limit:
\begin{align} \label{3.35}
\begin{split}
\Gamma_{k=0} [\,\overline{f}; \chib] 
& = 
\Gamma [\,\overline{f}; \chib] 
\\
\overline{\Gamma}_{k=0} [\phi]
& =
\overline{\Gamma} \,[\phi].
\end{split}
\end{align}
Hence $\Gamma_{k \to 0}$ and $\overline{\Gamma}_{k \to 0}$ satisfy the relations \eqref{3.14} and \eqref{3.15}, respectively. They entail that $\Gamma_{k=0} [\,\overline{f}; \chib]$ actually depends on the sum $\chib + \overline{f}$ only. This is \textit{not} true for $k \neq 0$, the reason being that in general $\Delta_k S [f; \chib]$ depends on $f$ and $\chib$ separately, not only on their sum. In the opposite limit $k \to \infty$, $\Gamma_{k} [\,\overline{f}; \chib]$ approaches $S [\chib + \overline{f}\,]$ plus a computable correction term, see \cite{liouv} for a detailed discussion of this point.
%
%
%
%
%
%
\section{The CREH Truncation}\label{s4}
\subsection{The Ansatz for $\boldsymbol{\Gamma_k}$}\label{s4.1}
In this section we specialize the as to yet exact flow equation \eqref{3.33} for the ``CREH truncation''\footnote{For a different approach to the quantization of conformal fluctuations see \cite{narpad}.}. It involves two approximations:
\begin{enumerate}
\item The usual Einstein--Hilbert truncation.
\item The conformal reduction: only the conformal factor is quantized while all other degrees of freedom contained in the metric as well as the Faddeev--Popov ghost fields are neglected.
\end{enumerate}
To make the presentation as transparent as possible we specialize for $d=4$ in the following. The key formulas for $d$ dimensions are summarized in Appendix \ref{AppB}.

The truncation ansatz for $\Gamma_{k} [\,\overline{f}; \chib]$ is given by the reduced functional $S_{\text{EH}} [\chib + \overline{f}\,]$ from eq.\ \eqref{2.7} with a $k$-dependent Newton constant $G_k$ and cosmological constant $\Lambda_k$:
\begin{align} \label{4.1}
\begin{split}
\Gamma_{k} [\,\overline{f}; \chib]
& =
- \frac{3}{4 \pi \, G_k} \,
\int \!\! \mathrm{d}^4 x~\sqrt{\widehat{g}\,} \,
\Big\{ - \tfrac{1}{2} \, \left( \chib + \overline{f}\, \right)
\, \widehat{\Box} \, \left( \chib + \overline{f}\, \right)
\\
& \phantom{{==}- \frac{3}{4 \pi \, G_k} \,
\int \!\! \mathrm{d}^4 x~\sqrt{\widehat{g}\,} \,
\Big\{}
+ \tfrac{1}{12} \, \widehat{R} \, \left( \chib + \overline{f}\, \right)^2
- \tfrac{1}{6} \, \Lambda_k \, \left( \chib + \overline{f}\, \right)^4 
\Big\}.
\end{split}
\end{align}
Here $\chib$ and $\overline{f}$ are still arbitrary functions of $x$. Parametrically the average action also depends on the, equally arbitrary, reference metric $\widehat{g}_{\mu \nu}$ with Ricci scalar $\widehat{R}$ and Laplace--Beltrami operator $\widehat{\Box}$. For this action the Hessian \eqref{3.34} has the form
$\langle x \lvert \Gamma_k^{(2)} \rvert y \rangle =
\Gamma_k^{(2)} \, \delta^4 (x-y) / \sqrt{\widehat{g} (x)\,}$ where $\Gamma_k^{(2)}$ is to be interpreted as a differential operator acting on $x$; it reads
\begin{align} \label{4.2}
\Gamma_{k}^{(2)} [\,\overline{f}; \chib]
& =
- \frac{3}{4 \pi \, G_k} \,
\Big\{ - \widehat{\Box}_x
+ \tfrac{1}{6} \, \widehat{R} (x)
- 2\, \Lambda_k \, \bigl( \chib (x) + \overline{f} (x) \bigr)^2 
\Big\}.
\end{align}
We shall come back to this operator shortly.
\subsection{The projected RG Equations}\label{s4.2}
In order to determine the $\beta$-functions for the running Newton constant $G_k$ and cosmological constant $\Lambda_k$ we proceed as follows. The first step consists in inserting the ansatz into the flow equation, both on its LHS, where we get $k$-derivatives of $G_k$ and $\Lambda_k$, and on its RHS where we are left with the problem of calculating a functional trace involving $\Gamma_k^{(2)}$. It is sufficient to compute this trace in a derivative expansion which retains only those terms which are also present on the LHS of the flow equation, namely those proportional to the monomials $\phi \, \widehat{\Box} \, \phi$, $\widehat{R} \, \phi^2$, and $\phi^4$ where $\phi \equiv \chib + \overline{f}$. If we then equate the coefficients of equal monomials on the LHS and RHS we find the desired RG equations of $G_k$ and $\Lambda_k$.

Without loosing information this calculation can be performed with a homogeneous background field: $\chib (x) = const \equiv \chib$.
The following two calculations are necessary then in order to ``project out'' the three monomials of interest:

\noindent\textbf{(i)} Evaluation of the functional trace for a flat metric $\widehat{g}_{\mu \nu}=\delta_{\mu \nu}$ and a non-zero, non-constant field $\overline{f} (x)$. Only the term $\overline{f} \, \widehat{\Box} \, \overline{f}$ must be retained. Comparing it to the relevant term of the LHS,
\begin{align} \label{4.3}
k \partial_k \, \Gamma_{k} [\,\overline{f}; \chib]
& =
+ \frac{3}{4 \pi} \, k \partial_k \, \left( \frac{1}{G_k} \right) \,
\int \!\! \mathrm{d}^4 x~\tfrac{1}{2} \, \overline{f} \, \widehat{\Box} \, \overline{f} + \cdots
\end{align}
yields the $\beta$-function of $G_k$.

\noindent\textbf{(ii)} Evaluation of the functional trace for $\overline{f} \equiv 0$ and $\widehat{g}_{\mu \nu}$ arbitrary whereby only the monomials $\chib^4$ and $\widehat{R}\, \chib^2$ are retained. Comparison with the corresponding terms on the LHS,
\begin{align} \label{4.4}
k \partial_k \, \Gamma_{k} [0; \chib]
& =
- \frac{3}{4 \pi} \, 
\int \!\! \mathrm{d}^4 x~\sqrt{\widehat{g}\,} \,
\left\{ \tfrac{1}{12} \, k \partial_k \, \left( \frac{1}{G_k} \right) \, \widehat{R} \, \chib^2
- \tfrac{1}{6} \, k \partial_k \, \left( \frac{\Lambda_k}{G_k} \right) \, \chib^4 + \cdots \right\}
\end{align}
allows for the computation of $\partial_k \left( \Lambda_k / G_k \right)$ and an alternative determination of $\partial_k G_k$.

Since both the $\phi \, \widehat{\Box} \, \phi$ and the $\widehat{R} \, \phi^2$ term appear with the same prefactor $1 / G_k$ we can derive a $\beta$-function for $G_k$ from either of them. We do not expect these $\beta$-functions to be exactly equal, but if our approximation makes sense they should be similar at least.

Before we can  embark on these calculations we must address the question of how $\mathcal{R}_k$ is to be adjusted. The IR cutoff at $k$ must be imposed on the spectrum of $\overline{\Box}$, not that of $\widehat{\Box}$. Since $\chib = const$ in the case at hand, the two operators are related by
\begin{align} \label{4.5}
\widehat{\Box} & = \chib^2 \, \overline{\Box}
\end{align}
so that we may reexpress $\Gamma_k^{(2)}$ as
\begin{align} \label{4.6}
\Gamma_{k}^{(2)} [\,\overline{f}; \chib]
& =
- \frac{3}{4 \pi \, G_k} \,
\Big\{ - \chib^2 \, \overline{\Box}
+ \tfrac{1}{6} \, \widehat{R}
- 2\, \Lambda_k \, \left( \chib + \overline{f}\, \right)^2 
\Big\}.
\end{align}
Now we define $\mathcal{R}_k$ in such a way that it leads to the replacement \eqref{3.26} when added to $\Gamma_k^{(2)}$:
\begin{align} \label{4.7}
\begin{split}
& \Gamma_{k}^{(2)} [\,\overline{f}; \chib] + \mathcal{R}_k [\chib]
\\
& \phantom{{=}} =
- \frac{3}{4 \pi \, G_k} \,
\Big\{ \chib^2 \, \bigl[ - \overline{\Box}
+ k^2 \, R^{(0)} (- \overline{\Box} / k^2) \bigr]
+ \tfrac{1}{6} \, \widehat{R}
- 2\, \Lambda_k \, \left( \chib + \overline{f}\, \right)^2 
\Big\}.
\end{split}
\end{align}
As a consequence, the cutoff operator has an explicit dependence on the background field:
\begin{align} \label{4.8}
\begin{split}
\mathcal{R}_k [\chib]
& =
- \frac{3}{4 \pi \, G_k} \, \chib^2 \, k^2 \, 
R^{(0)} \bigl(- \tfrac{\overline{\Box}\,}{k^2\,} \bigr)
\\
& =
- \frac{3}{4 \pi \, G_k} \, \chib^2 \, k^2 \, 
R^{(0)} \bigl(- \tfrac{\widehat{\Box}\,}{\chib^2 \, k^2\,} \bigr)
\end{split}
\end{align}
The two factors of $\chib^2$ appearing in the second line of \eqref{4.8} are the crucial difference between our treatment of the conformal factor and a standard scalar. If, instead of \eqref{3.26}, we had applied the ``substitution rule'' \eqref{3.27} they would have been absent.

Upon inserting the above $\mathcal{R}_k$ and reexpressing $\overline{\Box}$ as $\widehat{\Box} / \chib^2$ the flow equation assumes the form
\begin{align} \label{4.9}
\begin{split}
& k \partial_k \, \Gamma_{k} [\,\overline{f}; \chib]
\\
& \phantom{{=}} = \chib^2 \, k^2 \,
\tr \Biggl[
\left\{
\left( 1 - \tfrac{1}{2} \, \etan \right) \,
R^{(0)} \bigl( - \tfrac{\widehat{\Box}\,}{\chib^2 \, k^2\,} \bigr)
- \Bigl( - \frac{\widehat{\Box}}{\chib^2 \, k^2\,} \Bigr) \,
{R^{(0)}}' \bigl( - \tfrac{\widehat{\Box}\,}{\chib^2 \, k^2\,} \bigr)
\right\}
\\
& \phantom{{===}\chib^2 \, k^2 \,\tr \Biggl[} \times
\biggl( - \widehat{\Box} + \tfrac{1}{6} \, \widehat{R}
+ \chib^2 \, k^2 \, R^{(0)} \bigl( - \tfrac{\widehat{\Box}\,}{\chib^2 \, k^2\,} \bigr)
- 2\, \Lambda_k \, \left( \chib + \overline{f}\, \right)^2 
\biggr)^{-1}
\Biggr].
\end{split}
\end{align}
In evaluating the derivative $\partial_k \mathcal{R}_k$ we encountered the anomalous dimension $\etan$, defined in the same way as in \cite{mr}:
\begin{align} \label{4.10}
\etan & \equiv + k \partial_k \, \ln G_k.
\end{align}

Note that in eq.\ \eqref{4.9} the overall minus sign of $\mathcal{R}_k$, and hence $k \partial_k \mathcal{R}_k$, got canceled against the overall minus sign of $\Gamma_k^{(2)} + \mathcal{R}_k$ in \eqref{4.7}. This is the step where, within the present setting, the transition from the ``original'' to the ``inverted'' picture has taken place. The factor $(\cdots)^{-1}$ under the trace of \eqref{4.9} is the propagator of a mode with positive kinetic, but negative potential energy. This is an example of the ``$\mathcal{Z}_k = z_k$ rule'' discussed in \cite{mr} and \cite{oliver2}.

The only specialization which entered eq.\ \eqref{4.9} is $\chib = const$; the reference metric $\widehat{g}_{\mu \nu}$ and the fluctuation average $\overline{f}$ are still arbitrary. The equation \eqref{4.9} can serve as the starting point for the two calculations (i) and (ii) therefore. We start with the second one; it is analogous to the corresponding full Einstein--Hilbert calculation in ref.\ \cite{mr}, and a comparison will be instructive.

In (ii) we set $\overline{f} \equiv 0$ so that we are left with
\begin{align} \label{4.11}
k \partial_k \, \Gamma_{k} [0; \chib]
& = \chib^2 \, k^2 \,
\tr \left[
\mathcal{N} (- \widehat{\Box}) \,
\left( \mathcal{A} (- \widehat{\Box}) + \tfrac{1}{6} \, \widehat{R} (x)
\right)^{-1}
\right]
\end{align}
where we defined
\begin{align} \label{4.12}
\begin{split}
\mathcal{N} (- \widehat{\Box})
& \equiv
\left( 1 - \tfrac{1}{2} \, \etan \right) \,
R^{(0)} \bigl( - \tfrac{\widehat{\Box}\,}{\chib^2 \, k^2\,} \bigr)
- \Bigl( - \frac{\widehat{\Box}}{\chib^2 \, k^2\,} \Bigr) \,
{R^{(0)}}' \bigl( - \tfrac{\widehat{\Box}\,}{\chib^2 \, k^2\,} \bigr)
\\
\mathcal{A} (- \widehat{\Box})
& \equiv
- \widehat{\Box}
+ \chib^2 \, k^2 \, R^{(0)} \bigl( - \tfrac{\widehat{\Box}\,}{\chib^2 k^2\,} \bigr)
- 2\, \Lambda_k \, \chib^2 .
\end{split}
\end{align}
We need the trace on the RHS of \eqref{4.11} only to first order in $\widehat{R}$. Expanding the inverse operator in \eqref{4.11} and discarding higher order terms it remains to evaluate
\begin{align} \label{4.13}
\begin{split}
k \partial_k \, \Gamma_{k} [0; \chib]
& = \chib^2 \, k^2 \,
\tr \left[
\mathcal{N} (- \widehat{\Box}) \, 
\mathcal{A} (- \widehat{\Box})^{-1}
\right]
\\
& \phantom{{==}}
- \tfrac{1}{6} \, \chib^2 \, k^2 \,
\tr \left[ \widehat{R} (x) \, 
\mathcal{N} (- \widehat{\Box}) \, 
\mathcal{A} (- \widehat{\Box})^{-2}
\right].
\end{split}
\end{align}
At this point we assume that $\widehat{g}_{\mu \nu}$ is a metric on a maximally symmetric space\footnote{One could derive the beta-functions also without this assumption, with an identical result.}. Then $\widehat{R}$ is $x$-independent and can be pulled out of the second trace in \eqref{4.13}. The traces are then easily evaluated using the same expansion formula as in \cite{mr}:
\begin{align} \label{4.14}
\begin{split}
& \tr \bigl[ W (- \widehat{\Box}) \bigr]
\\
& \phantom{{=}}=
(4 \pi)^{-d/2} \, 
\left\{
Q_{d/2} [W] \, \int \!\! \mathrm{d}^d x~\sqrt{\widehat{g}\,}
+ \tfrac{1}{6} \, Q_{d/2-1} [W] \, 
\int \!\! \mathrm{d}^d x~\sqrt{\widehat{g}\,} \, \widehat{R}
+ \mathcal{O} (\widehat{R}\,^2)
\right\}.
\end{split}
\end{align}
Here, for any function $W$, $Q_0 [W] = W(0)$, and if $n>0$,
\begin{align} \label{4.15}
Q_n [W] & = \frac{1}{\Gamma (n)} \,
\int \limits_0^\infty \!\! \mathrm{d}z~z^{n-1} \, W (z).
\end{align}
Applying \eqref{4.14} to \eqref{4.13} yields
\begin{align} \label{4.16}
\begin{split}
k \partial_k \, \Gamma_{k} [0; \chib]
& = \frac{1}{(4 \pi)^2\,} \, \chib^2 \, k^2 \,
\left[
\tfrac{1}{6} \, \Big\{ Q_1 [\mathcal{N} / \mathcal{A}]
- Q_2 [\mathcal{N} / \mathcal{A}^2] \Big\}
\,  \int \!\! \mathrm{d}^4 x~\sqrt{\widehat{g}\,} \, \widehat{R}
\right.
\\
& \phantom{{==}\frac{1}{(4 \pi)^2\,} \, \chib^2 k^2 \,
\biggl[} \left.
+ Q_2 [\mathcal{N} / \mathcal{A}]
\,  \int \!\! \mathrm{d}^4 x~\sqrt{\widehat{g}\,}
+ \cdots
\right]
\end{split}
\end{align}
From \eqref{4.15} we obtain for the $Q_n$'s:
\begin{align} \label{4.17}
\begin{split}
Q_1 [\mathcal{N} / \mathcal{A}]
& = 
\Phi^1_1 (-2 \Lambda_k / k^2) - 
\tfrac{1}{2} \, \etan \, \widetilde{\Phi}^1_1 (-2 \Lambda_k / k^2)
\\
Q_2 [\mathcal{N} / \mathcal{A}^2]
& = 
\Phi^2_2 (-2 \Lambda_k / k^2) - 
\tfrac{1}{2} \, \etan \, \widetilde{\Phi}^2_2 (-2 \Lambda_k / k^2)
\\
Q_2 [\mathcal{N} / \mathcal{A}]
& = \chib^2 \, k^2 \,
\left[
\Phi^1_2 (-2 \Lambda_k / k^2) - 
\tfrac{1}{2} \, \etan \, \widetilde{\Phi}^1_2 (-2 \Lambda_k / k^2)
\right].
\end{split}
\end{align}
These expressions are written down in terms of the same standard threshold functions as in the full Einstein--Hilbert calculation \cite{mr}:
\begin{align} \label{4.18}
\begin{split}
\Phi^p_n (w)
& =
\frac{1}{\Gamma (n)} \, 
\int \limits_0^\infty \!\! \mathrm{d} z~z^{n-1} \,
\frac{R^{(0)} (z) - z \, {R^{(0)}}' (z)}
{\left[ z + R^{(0)} (z) + w \right]^p\,}
\\
\widetilde{\Phi}^p_n (w)
& =
\frac{1}{\Gamma (n)} \, 
\int \limits_0^\infty \!\! \mathrm{d} z~z^{n-1} \,
\frac{R^{(0)} (z)}{\left[ z + R^{(0)} (z) + w \right]^p\,}.
\end{split}
\end{align}
Without any further expansion or approximation, after a suitable change of the variable of integration, $Q_1 [\mathcal{N} / \mathcal{A}]$ and $Q_2 [\mathcal{N} / \mathcal{A}^2]$ have turned out independent of $\chib$, while $Q_2 [\mathcal{N} / \mathcal{A}]$ is proportional to $\chib^2$. As a result, the only relevant monomials appearing on the RHS of \eqref{4.16} are
$\int \! \mathrm{d}^4 x\sqrt{\widehat{g}\,} \, \chib^2 \, \widehat{R}$ and $\int \! \mathrm{d}^4 x\sqrt{\widehat{g}\,} \, \chib^4$. These are exactly the same ones as on the LHS of the flow equation, written out in \eqref{4.4}. By equating the coefficients of the first invariant we can read off the equation for $\partial_k G_k$; likewise the second yields $\partial_k (\Lambda_k / G_k)$, and upon using the information from the first, $\partial_k \Lambda_k$.

We write down this coupled system of two ordinary differential equations in terms of the dimensionless couplings
\begin{align} \label{4.19}
g_k \equiv k^2 \, G_k, \qquad 
\lambda_k \equiv \Lambda_k / k^2.
\end{align}
This choice of variables makes the system autonomous:
\begin{align}
\label{4.20}
k \partial_k \, g_k & = \beta_g (g_k, \lambda_k)
= \bigl[ 2 + \etan (g_k, \lambda_k) \bigr] \, g_k
\\
\label{4.21}
k \partial_k \, \lambda_k & = \beta_\lambda (g_k, \lambda_k).
\end{align}
The results for the anomalous dimension and the $\beta$-function of the dimensionless cosmological constant read explicitly
\begin{align} \label{4.22}
\etan^{\text{(pot)}} (g_k, \lambda_k)
& =
\frac{g_k \, B_1 (\lambda_k)}{1 - g_k \, B_2 (\lambda_k)}
\end{align}
and
\begin{align} \label{4.23}
\beta_\lambda (g_k, \lambda_k)
& =
- \left( 2 - \etan \right) \, \lambda_k
+ \frac{g_k}{2 \pi} \,
\left[ \Phi^1_2 (-2 \lambda_k) - 
\tfrac{1}{2} \, \etan \, \widetilde{\Phi}^1_2 (-2 \lambda_k)
\right].
\end{align}
The anomalous dimension involves the functions
\begin{align} \label{4.24}
\begin{split}
B_1 (\lambda_k)
& \equiv
\frac{1}{6 \pi} \, \left[
\Phi^1_1 (-2 \lambda_k) - \Phi^2_2 (-2 \lambda_k)
\right]
\\
B_2 (\lambda_k)
& \equiv
- \frac{1}{12 \pi} \, \left[
\widetilde{\Phi}^1_1 (-2 \lambda_k) - 
\widetilde{\Phi}^2_2 (-2 \lambda_k)
\right]
\end{split}
\end{align}
In eq.\ \eqref{4.22} we wrote $\etan^{\text{(pot)}}$ for $\etan$ in order to indicate that this expression for the anomalous dimension is derived from the potential (``pot'') term $\propto \widehat{R} \, \phi^2$. As we mentioned already, it can also be determined from the kinetic term $\propto (\partial_\mu \phi)^2$; the latter quantity will be denoted $\etan^{\text{(kin)}}$ later on.

Let us pause here for a moment to appreciate the similarity of the above ``scalar'' RG equations and those in the full Einstein--Hilbert truncation. The reader should compare the above equations \eqref{4.22}, $\cdots$, \eqref{4.24} to their counterparts in Section 4 of ref.\ \cite{mr}. In particular the structure \eqref{4.22} of the anomalous dimension is the same as in the full calculation. The $g_k \, B_2$-term in the denominator of \eqref{4.22} stems from the $k$-derivative of the $1 / G_k$-prefactor of the $\mathcal{R}_k$-operator \eqref{4.8}. It renders the result for $\etan$ manifestly nonperturbative in the sense that it sums up arbitrarily high orders of the coupling constant $g_k$.

The other calculation (i) which extracts $\etan \equiv \etan^{\text{(kin)}}$ from the kinetic term is performed in Appendix \ref{AppA}. The resulting RG equations are of the form \eqref{4.20}, \eqref{4.21} again, with the same $\beta_\lambda$, and with
\begin{align} \label{4.25}
\etan^{\text{(kin)}} (g_k, \lambda_k)
& =
- \frac{8}{3 \pi} \, g_k \, \lambda_k^2 \,\,
\frac{\widehat{\Sigma}_4 (-2 \lambda_k)}
{1 + \frac{4}{3 \pi} \, g_k \, \lambda_k^2 \, \,
\widetilde{\Sigma}_4 (-2 \lambda_k)}.
\end{align}
Here $\widehat{\Sigma}_4$ and $\widetilde{\Sigma}_4$ are special cases of two families of standard integrals containing the shape function $R^{(0)}$; the first one is defined by
\begin{align} \label{4.26}
\widehat{\Sigma}_d (w)
& \equiv
\frac{\mathrm{d}}{\mathrm{d} w} \, 
\Bigl( w^{(6-d)/2} \, \Sigma_d (w) \Bigr)
\end{align}
with
\begin{align} \label{4.27}
\Sigma_d (w) 
& \equiv
\frac{1}{d} \, \int \limits_0^\infty \!\! \mathrm{d} z~z^{d/2} \,
\frac{[ 1 + {R^{(0)}}' (z) ]^2}{[ z + R^{(0)} (z) + w ]^4\,}.
\end{align}
Here, as always, a prime denotes the derivative with respect to the argument. The second family is defined as
\begin{align} \label{4.28}
\widetilde{\Sigma}_d (w) 
& \equiv
\frac{2}{d} \, \int \limits_0^\infty \!\! \mathrm{d} z~z^{d/2} \,
\left\{
\frac{[ 1 + {R^{(0)}}' (z) ] \, {R^{(0)}}' (z)}
{[ z + R^{(0)} (z) + w ]^4\,}
- 2 \, \frac{[ 1 + {R^{(0)}}' (z) ]^2 \, R^{(0)} (z)}
{[ z + R^{(0)} (z) + w ]^5\,}
\right\}.
\end{align}

In the rest of this paper we shall analyze the two variants of the above RG equations in detail and compare their physical contents to that of the full Einstein--Hilbert truncation.
\subsection{The Beta Functions}\label{s4.3}
The ``shape'' of the cutoff, i.\,e.\ the precise form of the interpolation between the ``long--wavelength'' and the ``short--wavelength'' regime is governed by the shape function $R^{(0)}$. It is essentially arbitrary, except that it has to satisfy the boundary conditions $R^{(0)} (0) =1$, $R^{(0)} (\infty) =0$. At the level of observable quantities derived from the effective average action the $R^{(0)}$-dependence must cancel out (at least approximately). In the following we shall mostly employ the ``optimized'' shape function of ref.\ \cite{opt} since it allows for an analytic evaluation of the various integrals encountered above.

The optimized shape function is given by \cite{opt}
\begin{align} \label{4.29}
R^{(0)} (z) = \left( 1-z \right) \, \theta (1-z).
\end{align}
It has the following properties which help in simplifying the threshold functions:
\begin{gather} \label{4.30}
{R^{(0)}}' (z) = - \theta (1-z),
\qquad
{R^{(0)}}'' (z) = \delta (1-z)
\\ \nonumber
z + R^{(0)} (z) = \theta (1-z) + z \, \theta (z-1)
\\ \nonumber
R^{(0)} (z) - z \, {R^{(0)}}' (z) = \theta (1-z)
\\ \nonumber
\left[ 1 + {R^{(0)}}' (z) \right]^n = \theta (z-1),
\quad \forall n=1,2,3,\cdots
\\ \nonumber
\left[ 1 + {R^{(0)}}' (z) \right]^n \, \Bigl[ R^{(0)} (z) \Bigr]^m
= 0,
\quad \forall n,m=1,2,3,\cdots
\end{gather}
Even though $R^{(0)}$ is not smooth, all threshold functions are well behaved and no undefined distributions occur. For the $\Phi$-integrals one obtains \cite{litimgrav}
\begin{align} \label{4.31}
\begin{split}
\Phi^p_n (w) & = 
\frac{1}{\Gamma (n+1)} \, \frac{1}{\left( 1+w \right)^p\,}
\\
\widetilde{\Phi}^p_n (w) & = 
\frac{1}{\Gamma (n+2)} \, \frac{1}{\left( 1+w \right)^p\,}.
\end{split}
\end{align}
The integral \eqref{4.27} reduces to
\begin{align} \label{4.32}
\Sigma_d (w) & =
\frac{1}{d} \, \int \limits_1^\infty \!\! \mathrm{d} z~
\frac{z^{d/2}}{\left( z+w \right)^4\,}
\end{align}
which implies (by rescaling $z \to z/w$):
\begin{align} \label{4.33}
\widehat{\Sigma}_d (w) & =
\frac{1}{d} \, \frac{w^{(4-d)/2}}{\left( 1+w \right)^4\,}.
\end{align}
Furthermore one finds that, for this particular example of a shape function,
\begin{align} \label{4.34}
\widetilde{\Sigma}_d (w) & = 0.
\end{align}
For other shape functions $\widetilde{\Sigma}_d$ will be non-zero in general.

Using the above formulae we can write down the RG equations in explicit form. For the anomalous dimension coming from the kinetic term we obtain
\begin{align} \label{4.35}
\etan^{\text{(kin)}} (g_k, \lambda_k)
& =
- \frac{2}{3 \pi} \, 
\frac{g_k \,\lambda_k^2}{\left( 1 - 2 \lambda_k \right)^4\,}.
\end{align}
Furthermore, $\etan^{\text{(pot)}}$ is given by \eqref{4.22} with the following $B$-functions:
\begin{subequations} \label{4.36}
\begin{align}
\label{4.36a}
B_1 (\lambda_k)
& =
\frac{1}{3 \pi} \, \left( \frac{1}{4} - \lambda_k \right) \,
\frac{1}{\left( 1 - 2 \lambda_k \right)^2\,}
\\
\label{4.36b}
B_2 (\lambda_k)
& =
- \frac{1}{12 \pi} \, \left( \frac{1}{3} - \lambda_k \right) \,
\frac{1}{\left( 1 - 2 \lambda_k \right)^2\,}.
\end{align}
\end{subequations}
For $\beta_\lambda$ one finds
\begin{align} \label{4.37}
\beta_\lambda (g_k, \lambda_k)
& =
- \left( 2-\etan \right) \, \lambda_k
+ \frac{g_k}{4 \pi} \, \left( 1 - \frac{1}{6} \, \etan \right) \,
\frac{1}{1 - 2 \lambda_k}
\end{align}
where either $\etan^{\text{(kin)}}$ or $\etan^{\text{(pot)}}$ is to be inserted for $\etan$.

The above expressions highlight another similarity of the CREH calculation and the one in the full Einstein--Hilbert truncation: the poles at (or near) $\lambda = 1/2$. For a discussion of the singularities of $\beta_g$ and $\beta_\lambda$ we must distinguish the ``kin'' and the ``pot'' case.

Using $\etan^{\text{(kin)}}$, it is obvious from \eqref{4.35} and \eqref{4.37} that $\etan^{\text{(kin)}} (g, \lambda)$ and $\beta_\lambda (g, \lambda)$ have poles at $\lambda = 1/2$ and are regular otherwise. More precisely, on the $(g,\lambda)$-plane there exists the following line along which the $\beta$-functions diverge and the flow is undefined: $\{(g, 1/2) \, \lvert \, - \infty < g < +\infty \}$. The physically relevant part of the parameter space will be the half plane to the left of this line ($\lambda < 1/2$), as in the full theory.

With $\etan^{\text{(pot)}}$ the boundary of the ``physical'' parameter space is given by a curve to the left of the $\lambda=1/2$--line. Along this line, $1-g\, B_2 (\lambda) =0$, so that $\etan^{\text{(pot)}}$ diverges there, $\lvert \etan^{\text{(pot)}} \rvert = \infty$. Parameterizing this curve as $g = g_\eta^{\text{(pot)}} (\lambda)$ we have explicitly
\begin{align} \label{4.38}
g_\eta^{\text{(pot)}} (\lambda)
& =
12 \pi \, \frac{\left( 1- 2\lambda\right)^2}{\lambda - 1/3}.
\end{align}
It is amusing to note that at $\lambda = 1/2$, i.\,e.\ beyond the actual boundary, $\etan^{\text{(pot)}}$ and $\beta_\lambda$ are finite; for any $g \neq 0$ one has
\begin{subequations} \label{4.39}
\begin{align}
\label{4.39a}
\lim_{\lambda \nearrow 1/2} \etan^{\text{(pot)}} (g, \lambda) & = 6
\\
\label{4.39b}
\lim_{\lambda \nearrow 1/2} 
\beta_\lambda^{\text{(pot)}} (g, \lambda) & = 2.
\end{align}
\end{subequations}

In either of the two cases, the existence of a boundary in 
$(g, \lambda)$-space entails that some of the RG trajectories terminate already at a finite value of $k$ when they run into the boundary line. Within the full Einstein--Hilbert truncation, the status of the singularities has been discussed in detail in the literature \cite{frank1,frank2,h3,litimgrav}.
They have been interpreted as a breakdown of the truncation in the infrared. The continuation to $k=0$ would presumably require a more general ansatz for $\Gamma_k$.
\subsection{Comparison with the Standard Scalar FRGE}\label{s4.4}
One might wonder how the RG equations for the conformal factor relate to those for a standard scalar \cite{avactrev}. The comparison reveals that both the structure of the equation and their solutions are quite different in the two cases. We shall see this in more detail in Section \ref{s5}. Here we only mention the most striking deviation.

Let us consider a RG trajectory in a regime where the anomalous dimension is small so that we may approximate $\etan =0$. (In the next section we shall see that there are indeed trajectories with $\etan \approx 0$ over a large range of scales.) Then \eqref{4.10} integrates to $G_k = const \equiv \overline{G}$, and the equation for $\lambda_k$ involves the correspondingly simplified $\beta$-function \eqref{4.37}, with $g_k \equiv \overline{G} \, k^2$. In terms of the ordinary, dimensionful cosmological constant $\Lambda_k \equiv k^2 \, \lambda_k$ this RG equation reads
\begin{align} \label{4.40}
k \partial_k \, \Lambda_k
& =
\frac{\overline{G}}{4 \pi} \, \frac{k^6}{k^2 - 2 \Lambda_k}.
\end{align}
In particular, when $\Lambda_k \ll k^2$ it simplifies to
\begin{align} \label{4.41}
k \partial_k \, \Lambda_k
& =
\frac{1}{4 \pi} \, \overline{G} \, k^4.
\end{align}
Obviously \textit{the RG equations of the CREH truncation imply a quartic running of the cosmological constant as long as $\Lambda$ is small and $G$ is approximately constant.}

With quantum gravity in the back of our mind this result is no surprise. It is exactly what one finds in the full Einstein--Hilbert truncation \cite{mr}, except for the prefactor of $\overline{G} \, k^4$ which is anyhow non-universal. In fact, the $k^4$-running \eqref{4.41} is what all methods for summing zero--point energies would agree upon. In particular it can be seen as a reflection of the well known quartic divergences which appear in all Feynman diagram calculations (and are usually ``renormalized away''). So there can be no doubt that \eqref{4.41} is the physically correct answer for the regime considered.

On the other hand, from the scalar field perspective, the quartic running \textit{is} a surprise. In the CREH ansatz for $\Gamma_k$ the cosmological constant $\Lambda_k$ plays the role of a $\phi^4$-coupling constant which behaves as $\Lambda_k \propto k^4$ here. This very strong scale dependence has to be contrasted with the much weaker, merely logarithmic $k$-dependence one finds in an ordinary scalar theory on a $4$-dimensional flat spacetime (provided $k$ is above all mass thresholds, if any).

The origin of this significant difference in the RG running of the $\phi^4$-coupling, proportional to $\ln (k)$ for a standard scalar and $\propto k^4$ for the conformal factor, is clear: The conformal factor determines the proper scale of the cutoff, while a scalar matter field does not.
When we constructed the operator $\mathcal{R}_k$ in Subsection \ref{s4.2} we explained how the special status of the conformal factor comes into play. We saw that if the coarse graining scale is to be given a physical meaning, $k$ should be a cutoff in the spectrum of the \textit{background field dependent} operator $\overline{\Box}$, and this led to the substitution rule \eqref{3.26}.

Using instead the alternative rule for ordinary scalars, eq.\ \eqref{3.27}, our above calculation of the $\beta$-functions would have gone through almost unaltered up to eq.\ \eqref{4.13}. The only changes are: (a) The factors $\chib^2$ in front of the two traces on the RHS of \eqref{4.13} are absent, and (b), the operators $\mathcal{N}$ and $\mathcal{A}$ are replaced by
\begin{subequations} \label{4.42}
\begin{align}
\label{4.42a}
\mathcal{N}^{\, \text{standard}}
& =
\left( 1 - \tfrac{1}{2} \, \etan \right) \,
R^{(0)} ( - \widehat{\Box} / k^2 )
- \left( - \widehat{\Box} / k^2 \right) \,
{R^{(0)}}' ( - \widehat{\Box} / k^2 )
\\
\label{4.42b}
\mathcal{A}^{\, \text{standard}}
& =
- \widehat{\Box}
+ k^2 \, R^{(0)} ( - \widehat{\Box} / k^2 )
- 2\, \Lambda_k \, \chib^2 .
\end{align}
\end{subequations}
All $\chib$'s have disappeared, except the expected one coming from the second derivative of the interaction term $\propto \Lambda_k \, \left( \chib + f \right)^4$. It is clear that if one performs the calculations (i) and (ii) on the basis of the ``standard'' version of eq.\ \eqref{4.13} will be quite different. As the extra powers of $\chib^2$ are missing now, the pattern of terms to be equated on the two sides of the FRGE gets shifted correspondingly.

These remarks explain how, in a concrete calculation, ``background independence'' and the special status of the metric unavoidably lead to RG equations different from those of a scalar matter field.

In the construction of the exact gravitational average action in \cite{mr} where all degrees of freedom carried by the metric are quantized ``background independence'' and this special status have been taken care of. There it is the full background metric $\overline{g}_{\mu\nu}$, the generalization of $\chib^2$ here, which enters $\mathcal{R}_k$ and sets the scale of $k$. We emphasize that in \cite{mr} the use of the background field method serves two \textit{conceptually completely independent} purposes: (1) It allows for the definition of a physical (proper) coarse graining scale, as discussed in this paper, and (2), using an appropriate gauge fixing term, it leads to an average action $\Gamma_k [g_{\mu \nu}, \overline{g}_{\mu \nu}]$ which is a diffeomorphism invariant functional of its arguments. The latter property has no analog within the present setting.
%
%
%
\section{Asymptotic Safety in the CREH Truncation}\label{s5}
In this section we analyze the physical contents of the RG flow in the CREH truncation, being particularly interested in the asymptotic safety issue.
\subsection{Antiscreening}\label{s5.1}
From the definition \eqref{4.10} it follows that the RG running of the dimensionful Newton constant is given by
\begin{align} \label{5.1}
k \partial_k \, G_k & = \etan \, G_k.
\end{align}
If $\etan >0$, Newton's constant increases with increasing mass scale $k$, while it decreases if $\etan <0$. In analogy with gauge theory one refers to the first case as ``screening'', the second as ``antiscreening''. In the full\footnote{Here and in the following ``full calculation'' always refers to the complete calculation within the Einstein--Hilbert truncation in ref.\ \cite{mr}.} calculation, $\etan$ was of the antiscreening type in the entire physical part of the $(g, \lambda)$-plane.

If we determine \textbf{$\boldsymbol{\etan}$ from the kinetic term}, the corresponding CREH result is given by \eqref{4.35}. We observe that $\etan^{\text{(kin)}}$ is negative for any value of $g>0$ and $\lambda$. This corresponds to the antiscreening case: Newton's constant decreases at high energies. So the remarkable result is that the quantization of the conformal factor alone is already sufficient to obtain gravitational antiscreening. The spin-2 character of the metric field seems not essential and its selfinteractions (vertices) coming from $\int \! \mathrm{d}^4 x \sqrt{g\,} \, R$ seem not to play the dominant role. The only selfinteraction taken into account by the CREH calculation is the $\phi^4$-term stemming from $\Lambda \, \int \!\mathrm{d}^4 x \sqrt{g\,}$. However, in a sense the antiscreening within CREH is weaker than in the full theory. For example, close to the origin $g=\lambda=0$, the anomalous dimension in the full theory behaves as
\begin{align} \label{5.2}
\etan^{\text{(full)}} & = - 2 \, \omega \, g + \cdots
\end{align}
where $\omega$ is a positive \textit{constant}. For $\etan^{\text{(kin)}}$ the counterpart of $\omega$ is not constant but rather decreases $\propto \lambda^2$ when we approach the origin. A behavior of this sort was to be expected of course since the entire $\phi$-selfinteraction is due to the cosmological constant.

This trend is even more pronounced if we take the \textbf{$\boldsymbol{\etan}$ from the potential}. The corresponding expression is given in \eqref{4.22}. Its denominator is positive in the ``physical'' part of the $(g, \lambda)$-plane, so that it is $B_1 (\lambda)$ that decides about the sign of $\etan^{\text{(pot)}}$. Looking at \eqref{4.36a} we see that, if $g>0$,
\begin{align} \label{5.3}
\begin{split}
\etan^{\text{(pot)}} & \leq 0,
\quad \text{if } \lambda \geq 1/4, \\
\etan^{\text{(pot)}} & >0,
\quad \text{if } \lambda < 1/4.
\end{split}
\end{align}
The anomalous dimension $\etan^{\text{(pot)}}$ vanishes along the line $\lambda = 1/4$.
\subsection{Fixed Points}\label{s5.2}
Next we search for fixed points of the system of differential equations \eqref{4.20}, \eqref{4.21}, i.\,e.\ points $(g_\ast, \lambda_\ast)$ such that $\beta_g (g_\ast, \lambda_\ast) = 0 = \beta_\lambda (g_\ast, \lambda_\ast)$.

From \eqref{4.35}, \eqref{4.36}, and \eqref{4.37} it is obvious that for either choice of $\etan$ the system has a fixed point at the origin, referred to as the Gaussian Fixed Point (GFP): $g_\ast^{\text{GFP}} = \lambda_\ast^{\text{GFP}} = 0$.

A Non-Gaussian Fixed Point (NGFP), if any, would satisfy the condition $\beta_g =0$ with non-zero values of $g_\ast$ or $\lambda_\ast$ such that $\etan (g_\ast, \lambda_\ast) = -2$. Upon inserting $\etan=-2$ into \eqref{4.37} the condition $\beta_\lambda =0$ assumes the simple form
\begin{align} \label{5.4}
g_\ast & = 12 \pi \, \lambda_\ast \, \left( 1 - 2 \lambda_\ast
\right).
\end{align}
The second condition for $g_\ast$ and $\lambda_\ast$ depends on the choice for $\etan$.

If we use the \textbf{$\boldsymbol{\etan}$ from the kinetic term} given by eq.\ \eqref{4.35} the condition \linebreak$\etan^{\text{(kin)}} (g_\ast, \lambda_\ast)= -2$ reads
\begin{align} \label{5.5}
g_\ast \, \frac{\lambda_\ast^2}{\left( 1 - 2 \lambda_\ast \right)^4\,}
& = 3 \pi.
\end{align}
The system of equations \eqref{5.4}, \eqref{5.5} is easily decoupled by inserting $g_\ast$ of \eqref{5.4} into \eqref{5.5}. Remarkably, one does indeed find a real solution:
\begin{subequations} \label{5.6}
\begin{align}
\label{5.6a}
\lambda_\ast 
& = \frac{1}{2} \, \frac{2^{1/3}}{\left( 1 + 2^{1/3} \right)}
\approx 0.279
\\
\label{5.6b}
g_\ast 
& = 6 \pi \, \frac{2^{1/3}}{\left( 1 + 2^{1/3} \right)^2\,}
\approx 4.650
\end{align}
\end{subequations}
The existence of this NGFP comes as a true surprise; it has no counterpart in ordinary $4$-dimensional $\phi^4$-theory.

If instead we use the \textbf{$\boldsymbol{\etan}$ from the potential} given by \eqref{4.22} with \eqref{4.36} the condition $\etan^{\text{(pot)}} (g_\ast, \lambda_\ast) = -2$ can be written as
\begin{align} \label{5.7}
g_\ast \, \left( \lambda_\ast - \tfrac{5}{18} \right)
& = 4 \pi \, \left( 1 - 2 \lambda_\ast \right)^2.
\end{align}
The coupled equations \eqref{5.7} and \eqref{5.4} can be solved analytically again and they, too, give rise to real and positive fixed point coordinates:
\begin{subequations} \label{5.8}
\begin{align}
\label{5.8a}
\lambda_\ast 
& = \frac{7}{36} \, \left( \sqrt{481/49\,} -1 \right)
\approx 0.415
\\
\label{5.8b}
g_\ast 
& = 12 \pi \, \lambda_\ast \, \left( 1 - 2 \lambda_\ast \right)
\approx 2.665
\end{align}
\end{subequations}

The individual values of $g_\ast$ and $\lambda_\ast$ as obtained from the two calculational schemes do not quite agree. However, this does not come unexpected. The mere coordinates of the fixed point are not directly related to anything observable and, in fact, are scheme dependent or ``non-universal''. On the other hand, the product $g_\ast \lambda_\ast$ has been argued \cite{oliver1,oliver2} to be universal and can be measured in principle. And indeed, the products of the numbers in \eqref{5.6} and in \eqref{5.8} agree almost perfectly within the precision one could reasonably expect:
\begin{align} \label{5.9}
(g_\ast \lambda_\ast)^{\text{(kin)}} \approx 1.296,
\qquad
(g_\ast \lambda_\ast)^{\text{(pot)}} \approx 1.106
\end{align}

It is also gratifying to see that according to both calculations the respective NGFP is always located within the physical part of the $(g, \lambda)$-plane. The $\lambda_\ast$ coming from $\etan^{\text{(kin)}}$ satisfies $\lambda_\ast < 1/2$, see \eqref{5.6a}, and its counterpart from $\etan^{\text{(pot)}}$ given in \eqref{5.8a} can be shown to lie to the left of the ($g>0$ branch of the) boundary line $\lambda \mapsto g_\eta^{\text{(pot)}} (\lambda)$ from eq.\ \eqref{4.38}.

It is straightforward to generalize the calculations for arbitrary dimensionalities $d$. The corresponding formulas are listed in Appendix \ref{AppB}, and the numerical results for $d$ between $3$ and $10$ are displayed in Table \ref{tab1}. 
\begin{table}
\begin{center}
\begin{tabular}{|r||c|c|c|c|c|c|}
\hline
$\boldsymbol{d}$ & \textbf{Trunc.} & $\boldsymbol{g_\ast}$ & $\boldsymbol{\lambda_\ast}$ 
& $\boldsymbol{\tau_d}$ & $\boldsymbol{\theta'}$ & $\boldsymbol{\theta''}$ \\
\hline
\hline
3 & full EH & 0.202139 & 0.0651806 & 0.00266329 & 1.11664 & 0.827598 \\
\hline
& CREH, pot & 0.172872 & 0.233092 & 0.00696588 & -3.54754 & 4.92795 \\
\hline
& CREH, kin & 0.391798 & 0.126945 & 0.0194868 & 2.04572 & 3.60445 \\
\hline
\hline
4 & full EH & 0.707321 & 0.193201 & 0.136655 & 1.4753 & 3.04321 \\
\hline
& CREH, pot & 2.6654 & 0.41477 & 1.10553 & 1.47122 & 9.30442 \\
\hline
& CREH, kin & 4.65005 & 0.278753 & 1.29622 & 4.0 & 6.1837 \\
\hline
\hline
5 & full EH & 2.85863 & 0.234757 & 0.472851 & 2.76008 & 5.12941 \\
\hline
& CREH, pot & 26.9696 & 0.557727 & 5.01577 & 5.81627 & 12.0556 \\
\hline
& CREH, kin & 42.3258 & 0.417188 & 5.06681 & 6.27681 & 8.6899 \\
\hline
\hline
6 & full EH & 13.8555 & 0.255477 & 0.950958 & 4.48592 & 7.07967 \\
\hline
& CREH, pot & 243.547 & 0.674559 & 10.5272 & 10.8493 & 14.3777 \\
\hline
& CREH, kin & 361.57 & 0.537523 & 10.221 & 8.81712 & 11.1844 \\
\hline
\hline
7 & full EH & 76.3589 & 0.269073 & 1.5241 & 6.51007 & 8.9431 \\
\hline
& CREH, pot & 2134.67 & 0.77282 & 16.5886 & 17.0223 & 15.9635 \\
\hline
& CREH, kin & 3069.3 & 0.641211 & 15.9154 & 11.591 & 13.6754 \\
\hline
\hline
8 & full EH & 464.662 & 0.279376 & 2.16389 & 8.78536 & 10.7446 \\
\hline
& CREH, pot & 18744.8 & 0.857143 & 22.7691 & 24.6444 & 16.0092 \\
\hline
& CREH, kin & 26451.9 & 0.730796 & 21.7745 & 14.5789 & 16.1597 \\
\hline
\hline
9 & full EH & 3066.23 & 0.287851 & 2.85326 & 11.2969 & 12.4932 \\
\hline
& CREH, pot & 167205.0 & 0.930559 & 28.9135 & 33.9881 & 12.4239 \\
\hline
& CREH, kin & 233516.0 & 0.808694 & 27.6432 & 17.7666 & 18.6307 \\
\hline
\hline
10 & full EH & 21673.5 & 0.295179 & 3.58153 & 14.044 & 14.1871 \\
\hline
& CREH, pot & 1.52489$ \cdot 10^6$ & 0.995177 & 34.9712 & & \\
\hline
& CREH, kin & 2.11943$ \cdot 10^6$ & 0.876935 & 33.4597 & 21.1433 & 21.0813 \\
\hline
\end{tabular}
\end{center}
\caption{}
\label{tab1}
\end{table}
In all dimensions considered a NGFP is found to exist, with $g_\ast > 0$ and $\lambda_\ast >0$. For each value of $d$, the table contains the results obtained within the full Einstein--Hilbert (EH) truncation as well as the ``pot'' and ``kin'' variants of the CREH truncation. It contains also the generalization of $g_\ast \lambda_\ast$, namely $\tau_d \equiv \lambda_\ast g_\ast^{2/(d-2)}$ which is the fixed point value of the dimensionless combination
$\Lambda_k G_k^{2/(d-2)} = \lambda_k g_k^{2/(d-2)}$. (Note that in $d$ dimensions $g_k = k^{d-2} \, G_k$ and $\lambda_k = k^{-2} \, \Lambda_k$.) It is impressive to see how well the ``pot'' and ``kin'' values of $\tau_d$ agree for any $d \geq 4$. As compared to the full Einstein--Hilbert result, the $\tau_d$-values are always larger by about a factor of 10.

We interpret this factor as indicating that the conformal factor is not the only degree of freedom driving the formation of a NGFP, but its contribution is typical in the sense that it leads to an RG flow which is qualitatively similar to the full one.
\subsection{Critical Exponents of the NGFP}\label{s5.3}
The properties of the RG flow on $(g, \lambda)$-space linearized about the NGFP are determined by the stability matrix
\begin{align} \label{5.10}
B & =
\begin{bmatrix}
\frac{\partial \beta_\lambda}{\partial \lambda} & 
\frac{\partial \beta_\lambda}{\partial g} \\
\frac{\partial \beta_g}{\partial \lambda} & 
\frac{\partial \beta_g}{\partial g}
\end{bmatrix}
\end{align}
evaluated at $(g_\ast, \lambda_\ast)$. Using the same notation as in \cite{oliver1,frank1} we write the corresponding eigenvalue problem as $B \, V = - \theta \, V$ and refer to the negative eigenvalues $\theta$ as the ``critical exponents''. In general $B$ is not expected to be symmetric.

Employing the \textbf{$\boldsymbol{\etan}$ from the kinetic term} the explicit form of the stability matrix, evaluated at the respective fixed point coordinates, is found to be
\begin{align} \label{5.11}
B & =
\begin{bmatrix}
-6 & \frac{1}{4 \pi} \, \left( 1 + 2^{1/3} \right) \\
- 48 \pi \, \frac{\left( 1 + 2^{4/3} \right)}{\left( 1 + 2^{1/3} 
\right)} & -2
\end{bmatrix}.
\end{align}
The resulting eigenvalues are non-zero and complex. The two critical exponents $\theta_{1,2} = \theta' \pm i \theta''$ form a complex conjugate pair with real and imaginary parts given by, respectively,
\begin{align} \label{5.12}
\begin{split}
\theta' & = 4
\\
\theta'' & = 2 \, \sqrt{2\,} \, \sqrt{1 + 3 \cdot 2^{1/3}\,}
\approx 6.1837
\end{split}
\end{align}
The positive real part indicates that the NGFP is UV attractive (attractive for $k \to \infty$) in both directions of the $(g, \lambda)$-plane. The non-vanishing imaginary part implies that near the NGFP the RG trajectories are spirals. This is exactly the same pattern as in the full Einstein--Hilbert truncation \cite{frank1}.

Using instead the \textbf{$\boldsymbol{\etan}$ from the potential} we find the same qualitative behavior, but the exponents are somewhat different:
\begin{align} \label{5.13}
\begin{split}
\theta' & \approx 1.471
\\
\theta'' & \approx 9.304
\end{split}
\end{align}

The discrepancy between \eqref{5.12} and \eqref{5.13} can serve as a rough measure for the accuracy of the calculation. First of all it is gratifying to see that both calculations lead to the same qualitative behavior: attractivity in both directions of parameter space, and a non-zero imaginary part. Numerically, the values for $\theta'$ and $\theta''$ probably can be trusted only within a factor of $2$ or so. In Table \ref{tab1} we display the critical exponents also for the other dimensions and compare them to the values in the full calculation. For $d>4$ the situation is essentially the same as in $d=4$: While the $\tau$-values of the two CREH versions agree very well, the agreement of the $\theta$'s is poorer. This is in line with an observation made in earlier calculations \cite{oliver1,oliver2,frank1,litimgrav,codello} namely that it is much easier to get a precise value for $g_\ast \lambda_\ast$ than for the critical exponents. In $d=3$ the CREH truncation seems insufficient to deduce the qualitative behavior reliably; the sign of $\theta'$ depends on whether $\etan^{\text{(pot)}}$ or $\etan^{\text{(kin)}}$ is used.

With regard to the numbers in Table \ref{tab1} it has to be emphasized, however, that even in an exact treatment of the conformally reduced theory the resulting critical exponents would have no reason to agree with those from full QEG which quantizes also the other degrees of freedom contained in the metric. The field contents of the two theories is different, and so one would expect them to belong to different universality classes, with different $\theta$'s.
\subsection{The Phase Portrait}\label{s5.4}
Finally we solve the coupled equations \eqref{4.20}, \eqref{4.21} numerically in order to obtain the phase portrait of the CREH flow. Using the anomalous dimension $\etan^{\text{(kin)}}$ we find the result displayed in Fig.\ \ref{fig1}. 
\begin{figure}
\renewcommand{\baselinestretch}{1.1}
\centering
\pgfuseimage{1a}\\
{\footnotesize (a)}
\\[1em]
\pgfuseimage{1b}\\
{\footnotesize (b)}
\caption{The figures show the RG flow on the $(g, \lambda)$-plane which is obtained from the CREH truncation with $\etan^{\text{(kin)}}$. The arrows point in the direction of decreasing $k$.}
\label{fig1}
\end{figure}
This flow diagram is strikingly similar to the corresponding diagram of the full Einstein--Hilbert truncation\footnote{See the diagram in Fig.\ 12 of ref.\ \cite{frank1}.}. The flow is dominated by the NGFP and the GFP at the origin, and we can distinguish three types of trajectories spiraling out of the NGFP. They are heading for negative, vanishing, and positive cosmological constant, respectively, and correspond exactly to the Type Ia, IIa, and IIIa trajectories of the full flow \cite{frank1}. The trajectories of the CREH Types Ia and IIa extend down to $k=0$, those of Type IIIa terminate at a non-zero $k_{\text{term}}$ when they reach $\lambda=1/2$, exactly as in the full theory. It is also interesting to note that this approximation admits RG trajectories whose turning point is arbitrarily close to the GFP so that there emerges a long classical regime \cite{h3}. (Such trajectories might be relevant in cosmology \cite{h3,entropy}.) The only boundary of the physical domain of the CREH parameter space is the vertical line at $\lambda =1/2$. The basin of attraction of the NGFP consists of all points $(g, \lambda)$ in this domain with $g>0$, exactly as in the full EH truncation.

If we solve the RG equations with the second version of the anomalous dimension, $\etan^{\text{(pot)}}$, we obtain the phase portrait shown in Fig.\ \ref{fig2}a.
\begin{figure}
\renewcommand{\baselinestretch}{1.1}
\centering
\pgfuseimage{2a}\\
{\footnotesize (a)}
\\[1em]
\pgfuseimage{2b}\\
{\footnotesize (b)}
\caption{As in Figs.\ \ref{fig1}, but with $\etan^{\text{(pot)}}$. The fat line is the boundary of the physical parameter space on which $\etan^{\text{(pot)}}$ diverges. In Fig.\ \ref{fig2}b also the ``unphysical'' flow beyond the boundary is shown.}
\label{fig2}
\end{figure}
In the vicinity of the GFP and NGFP, respectively, the structure of the flow, again, is exactly the same as in the full theory. The only new feature here is that there exist trajectories which begin and end on the boundary of the physical part of $(g, \lambda)$-space which is given by eq.\ \eqref{4.38}. As a result, the basin of attraction is smaller than the full physical domain, and trajectories with a turning point close to the NGFP are outside this basin. Even though this feature is different from the full EH flow we see that the conformal factor drives the flow in the same direction as the full metric and is in this sense representative. It is, however, too weak to push the trajectories sufficiently strongly away from the hyperbolic shape they have in absence of any non-trivial RG effects \cite{h3}. In Fig.\ \ref{fig2}b also some ``unphysical'' trajectories to the right of the singular line are shown. Note that at the singularity the direction of the arrows is reversed.

The overall conclusion of this section is that the RG flow implied by the CREH approximation, at least in a neighborhood of the two fixed points, is qualitatively identical to that obtained with the full Einstein--Hilbert truncation. In particular both versions of $\etan$ agree on the existence of a NGFP with precisely the properties required for asymptotic safety.
%
%
%
\section{Conclusion}\label{s6}
In this paper we used the conformally reduced Einstein--Hilbert truncation in order to explore a property of gravitational RG flows which is completely general presumably, at least in the framework of the effective average action. Among all the fields we use to describe Nature the metric enjoys a special status since it fixes the proper value of any dimensionful physical quantity. When one applies the Kadanoff--Wilson interpretation of RG flows as a sequence of consecutive coarse graining steps to quantum gravity one would like to give an, at least approximate, physical meaning to the notion of a coarse graining scale. The effective average action $\Gamma_k [g_{\mu\nu}, \overline{g}_{\mu \nu}]$ meets this requirement by introducing $k$ as a cutoff in the spectrum of the covariant Laplacian pertaining to the background metric $\overline{g}_{\mu \nu}$. Hence the mass scale $k$ is ``proper'', in the sense of the ``cutoff modes'' \cite{jan1,jan2}, with respect to $\overline{g}_{\mu \nu}$. Therefore the mode suppression term $\Delta_k S$ and, as a result, the cutoff operator $\mathcal{R}_k$ depend on the background field in a non-trivial way even if a simple truncation is used which does not involve $\overline{g}_{\mu \nu}$ explicitly. The $\overline{g}_{\mu \nu}$-dependence of the cutoff has a strong impact on the resulting RG flow. We argued that to some extent asymptotic safety, the formation of a non-Gaussian fixed point, is an essentially ``kinematical'' phenomenon resulting from this special role played by the metric.

At the same time the use of the field-, i.\,e.\ the $\overline{g}_{\mu \nu}$-dependent cutoff operator is forced upon us by the requirement of ``background independence''. In standard matter field theory $\mathcal{R}_k$ involves the metric of the fixed spacetime manifold on which the theory is defined, Minkowski space, say. In quantum gravity we are aiming at a ``background independent'' quantization. This suggests that if a gravitational RG flow is to have any physical meaning it should be ``background independent'', too, that is, it should not involve any special metric in its construction. If $\mathcal{R}_k$ is constructed from one of the arguments of $\Gamma_k [g_{\mu \nu}, \overline{g}_{\mu \nu}\,]$ this is indeed the case. In the effective average action $\overline{g}_{\mu \nu}$ is used for this purpose because only then $\Delta_k S$ is quadratic in the fluctuation fields and the FRGE contains second field derivatives only.

We illustrated these issues by means of the CREH truncation which quantizes only one of the degrees of freedom contained in the metric, the conformal factor. If we ignore the special status of the metric we arrive at the $\phi^4$-theory with a negative quartic coupling which, according to Symanzik, is asymptotically free. If, instead, the metric itself is used to set the proper scale of $k$, then the RG flow is different and in particular a NGFP exists. It is quite remarkable that, at the qualitative level, this simple scalar--like theory has exactly the same flow diagram as the full Einstein--Hilbert truncation. It is therefore plausible to conjecture that the complicated selfinteractions of the helicity-2 modes, another feature that distinguishes the metric from matter fields, is possibly not at the heart of asymptotic safety in gravity. In the CREH approximation only the interactions due to the cosmological constant term $\Lambda \, \sqrt{g\,} \propto \Lambda \, \phi^4$ were retained.

The experience with the CREH truncation suggests that the RG flow due to the conformal factor alone might be typical of the full metric. Therefore the obvious next step is to apply the conformal reduction to more general truncations. In the companion paper \cite{creh2} we therefore analyze the local potential approximation for $\phi$ as a first step in this direction. These investigations might help in making contact with other approaches to quantum gravity, numerical simulations, for instance \cite{ajl1,ajl2,ajl34}.

\vspace{3\baselinestretch pt}
\noindent
Acknowledgement: M.R.\ would like to thank A.~Ashtekar, L.~Freidel, M.~Niedermaier,\linebreak R.~Percacci, C.~Rovelli, and T.~Thiemann for helpful discussions.
%
%
%
\pagebreak
\appendix
\noindent {\Large \textbf{APPENDICES}}
\section{Derivation of the Anomalous Dimension $\boldsymbol{\etan^{\text{(kin)}}}$}\label{AppA}
In this appendix we derive the anomalous dimension $\etan^{\text{(kin)}}$ given in eq.\ \eqref{4.25} which follows from the kinetic term in the truncation ansatz. Actually we shall derive a slightly more general result which covers also the case of the local potential approximation (LPA) that will be studied in the companion paper \cite{creh2}. We shall first look at a standard scalar flow equation \cite{tet} and then point out the differences in the gravitational case.

We try to solve the standard scalar FRGE
\begin{align} \label{a.1}
k \partial_k \, \Gamma_k [\phi]
& = 
\tfrac{1}{2} \, \tr \left[ \left( \Gamma_k^{(2)} + \mathcal{R}_k \right)^{-1} \, k \partial_k \, \mathcal{R}_k\right]
\end{align}
on $d$-dimensional flat Euclidean space. A still rather general LPA truncation would be
\begin{align} \label{a.2}
\Gamma_k [\phi]
& =
\int \!\! \dd^d x ~
\Big \{ \tfrac{1}{2} \, \mathbf{Z}_k \bigl( \phi (x) \bigr) \, 
\partial_\mu \phi \, \partial_\mu \phi + U_k \bigl( \phi (x) \bigr)
\Big \}
\end{align}
where $\mathbf{Z}_k (\cdot)$ and $U_k (\cdot)$ are two running functions of $\phi$. Here we shall be more modest, however, and retain only the arbitrary field dependence of the potential, while we fix
\begin{align} \label{a.3}
Z_k & \equiv \mathbf{Z}_k \bigl( \phi_1 (k) \bigr)
\end{align}
at some, possibly $k$-dependent, reference point $\phi = \phi_1 (k)$. Typically $\phi_1$ is taken to be the minimum of the potential $U_k$. (In the gravitational case this will not always be possible, however \cite{creh2}.) Applying the ''$\mathcal{Z}_k=Z_k$ rule'', the cutoff operator for this truncation is $\mathcal{R}_k =
Z_k \, k^2 \, R^{(0)} (-\Box / k^2)$. We would like to use the FRGE \eqref{a.1} in order to derive the beta-function of $Z_k$, or equivalently the anomalous dimension
\begin{align} \label{a.4}
\eta & \equiv - Z_k^{-1} \, k \p_k \, Z_k.
\end{align}
We project out the kinetic term from $\Gamma_k [\phi]$ by inserting the special field configuration $\phi (x) = \phi_1 + \varphi (x)$ and expanding up to second order in both $\varphi (x)$ and $\p_\mu$. Upon inserting the truncation into the exact flow equation we thus obtain, omitting irrelevant terms,
\begin{align} \label{a.5}
\begin{split}
& k \p_k \, Z_k \int \!\! \dd^d x ~
\tfrac{1}{2} \, \p_\mu \varphi (x) \, \p_\mu \varphi (x)
+ \dotsm
\\
& \phantom{{=}} =
k^2 \, \tr 
\biggl[ \Big \{ \left( 1 - \eta /2 \right) \, 
R^{(0)} \left( \h p\,^2 / k^2 \right)
- \left( \h p\,^2 / k^2 \right) \, {R^{(0)}}' \left( \h p\,^2 / k^2 \right)
\Big \}
\\
& \phantom{{===} k^2 \, \tr \biggl[}
\times \Bigl( \h p\,^2 + k^2 \, R^{(0)} \left( \h p\,^2 / k^2 \right)
+ U_k'' \bigl( \phi_1 + \varphi (\h x\,) \bigr) / Z_k \Bigr)^{-1} 
\biggr].
\end{split}
\end{align}
Here we employed the same Hilbert space notation as in elementary quantum mechanics, with operators $\h x_\mu$ and $\h p_\mu$ satisfying $[ \h x_\mu, \h p_\nu ] = i \delta_{\mu \nu}$. In the position representation we have $\h x_\mu = x_\mu$, $\h p_\mu = -i \p_\mu$, but clearly the trace of \eqref{a.5} can be computed in any representation. We now apply standard derivative expansion techniques \cite{fraser} in order to pull out the $\int (\p_\mu \varphi)^2$--contribution from the trace which then, by \eqref{a.5}, determines $k \p_k \, Z_k$. It will be convenient to introduce the functions
\begin{align}
\label{a.6}
P (p^2)
& \equiv
p^2 + k^2 \, R^{(0)} \left( p^2 / k^2 \right)
\\
\label{a.7}
H (p^2)
& \equiv
P (p^2) + \gamma
\\
\label{a.8}
N (p^2)
& \equiv
k \p_k \, P (p^2) - \eta \, \bigl( P (p^2) - p^2 \bigr).
\end{align}
We also abbreviate
\begin{align} \label{a.9}
& & & & 
\alpha
& \equiv
U_k''' (\phi_1) / Z_k,
& \gamma
& \equiv
U_k'' (\phi_1) / Z_k
& & & & 
\end{align}
where, as always, the prime denotes a derivative with respect to the argument. Thus \eqref{a.5} becomes, up to irrelevant terms,
\begin{align} \label{a.10}
\begin{split}
& k \p_k \, Z_k \int \!\! \dd^d x ~
\p_\mu \varphi (x) \, \p_\mu \varphi (x)
+ \dotsm
\\
& \phantom{{=}}=
\tr \Bigl[
N (\h p\,^2) \,
\bigl( H (\h p\,^2) + \alpha \, \varphi (\h x\,) \bigr)^{-1} \Bigr]
\equiv \alpha^2 \, \mathscr{S}.
\end{split}
\end{align}
In expanding $U_k'' (\phi_1 + \phi (\h x\,))$ it was enough to retain the term linear in $\varphi$ since the next term, $U_k''' (\phi_1) \, \varphi^2 (x)$, has two $\varphi$'s at the same $x$ and hence cannot contribute to the $\int (\p_\mu \varphi)^2$--invariant. If we expand the denominator in \eqref{a.10} and retain only the term quadratic in $\varphi$ we arrive at
\begin{align} \label{a.11}
\mathscr{S}
& =
\tr \Bigl(
N \, H ^{-2} \, \varphi \, H^{-1} \, \varphi \Bigr)
\end{align}
where we also exploited the cyclicity of the trace and $[N,H]=0$. Note that $N$ and $H$ contain only the operators $\h p_\mu$, while $\varphi$ contains only $\h x_\mu$. Eq.\ \eqref{a.11} is trivially equal to
\begin{align} \label{a.12}
\mathscr{S}
& =
\tr \Bigl(
N \, H^{-2} \, \left[ \varphi, 
H^{-1} \right] \, \varphi \Bigr)
+
\tr \Bigl( N \, H^{-3} \, \varphi^2 \Bigr).
\end{align}
We may ignore the second trace on the RHS of \eqref{a.12} since in the position representation it is easy to see that it cannot give rise to a term with derivatives acting on the $\varphi$'s. In order to evaluate the first trace we set $F (p^2) \equiv 1 / H (p^2)$ and apply the familiar commutator identity
\begin{align} \label{a.13}
\begin{split}
\bigl[ \varphi (\h x\,), F (\h p\,^2) \bigr]
& =
2 i \, F' (\h p\,^2) \, \h p_\mu \, \left( \p_\mu \varphi \right)
(\h x\,)
+ F' (\h p\,^2) \, \left( - \Box \varphi \right) (\h x\,)
\\
& \phantom{{==}}
+ 2 \, F'' (\h p\,^2) \, \h p_\mu \h p_\nu \,
\left( - \p_\mu \p_\nu \varphi \right) (\h x\,)
+ \mathcal{O} (\p^3 \varphi).
\end{split}
\end{align}
Upon inserting \eqref{a.13} into \eqref{a.12} all $\h x\,$'s appear to the right of all $\h p\,$'s so that in the $x$-representation
\begin{align} \label{a.14}
\begin{split}
\mathscr{S}
& =
\int \!\!\ \dd^d x~ \langle x \rvert \,
N \, H^{-2} \, F' (\h p\,^2) 
\, \left( - \Box \varphi \right) (\h x\,) \, \varphi (\h x\,)
\,\lvert x \rangle
\\
& \phantom{{==}}
+ 2 \, \int \!\!\ \dd^d x~ \langle x \rvert \,
N \, H^{-2} \, F'' (\h p\,^2) \, 
\h p_\mu \h p_\nu \, \left( - \p_\mu \p_\nu \varphi \right) (\h x\,) \,
\varphi (\h x\,)
\,\lvert x \rangle.
\end{split}
\end{align}
Now we can exploit that $\h x_\mu \lvert x \rangle = x_\mu \lvert x \rangle$:
\begin{align} \label{a.15}
\begin{split}
\mathscr{S}
& =
\int \!\!\ \dd^d x~ 
\Bigl( \langle x \rvert \,
N \, H^{-2} \, F' (\h p\,^2) 
\,\lvert x \rangle \, \delta_{\mu \nu}
\\
& \phantom{{==}\int \!\!\ \dd^d x~ \Bigl(}
+ 2 \, \langle x \rvert \,
N \, H^{-2} \, F'' (\h p\,^2) \, 
\h p_\mu \h p_\nu 
\,\lvert x \rangle
\Bigr) \, \p_\mu \varphi (x) \, \p_\nu \varphi (x).
\end{split}
\end{align}
The remaining matrix elements $\langle x \rvert \dotsm \lvert x \rangle$, by translation invariance, are actually $x$-independent and can easily be evaluated in momentum space. As $p_\mu p_\nu$ is equivalent to $p^2\, \delta_{\mu \nu}/d$ under the momentum integral both terms on the RHS of \eqref{a.15} are seen to contribute to the $\int \p_\mu \varphi \, \p_\mu \varphi$--invariant. We can now read off that
\begin{align} \label{a.16}
\begin{split}
k \p_k \, Z_k
& =
\tfrac{1}{2} \, \alpha^2 \, \int \!\! \frac{\dd^d p}{(2 \pi)^d\,}~
\frac{N (p^2)}{H (p^2)^2\,} \,
\Bigl( 2 \, F' (p^2) + \tfrac{4}{d} \, p^2 \, F'' (p^2)
\Bigr)
\\
& =
v_d \, \alpha^2 \, \int \limits_0^\infty \!\! \dd y~
\frac{N (y)}{H (y)^2\,} \,
\Bigl( 2 \, y^{d/2 -1} \, F' (y) + \tfrac{4}{d} \, y^{d/2} \, F'' (y)
\Bigr)
\\
& =
\frac{4 \, v_d \, \alpha^2}{d} \, \int \limits_0^\infty \!\! \dd y~ 
y^{d/2} \, \frac{H'(y)}{H (y)^2\,}
\frac{\dd}{\dd y} \left( \frac{N(y)}{H (y)^2\,} \right).
\end{split}
\end{align}
Here $y \equiv p^2$ and $v_d \equiv [2^{d+1} \, \pi^{d/2} \, \Gamma (d/2) ]^{-1}$. The last equality in \eqref{a.16} follows by noticing that $2 \, y^{d/2-1} \, F' (y) + \tfrac{4}{d} \, y^{d/2} \, F'' (y) = \tfrac{\dd}{\dd y} \, \left( \tfrac{4}{d} \, y^{d/2} \, F'(y) \right)$, using $F' = - H'/H^2$, and performing an integration by parts. Recalling the definitions of $\eta$ and $N$ eq.\ \eqref{a.16} has the structure
\begin{align} \label{a.17}
\eta
& =
- 2 \, v_d \, \alpha^2 \, Z_k^{-1} \, \bigl( I_1 - \eta \, I_2 \bigr)
\end{align}
with the integrals
\begin{align} 
\label{a.18}
I_1
& \equiv
\frac{2}{d} \, \int \limits_0^\infty \!\! \dd y~ 
y^{d/2} \, \frac{P'(y)}{H (y)^2\,} \,
\frac{\dd}{\dd y} \frac{k \p_k \, P (y)}{H (y)^2\,}
\\
\label{a.19}
I_2 
& \equiv
\frac{2}{d} \, \int \limits_0^\infty \!\! \dd y~ 
y^{d/2} \, \frac{P'(y)}{H (y)^2\,} \,
\frac{\dd}{\dd y} \frac{P (y) - y}{H (y)^2\,}.
\end{align}

In order to further evaluate $I_1$ it is convenient to introduce the modified scale derivative $\w \p_k$ which, by definition, acts only on the explicit $k$-dependence in $k^2 \, R^{(0)} (y/k^2)$, but not on the implicit one in $Z_k$, $U_k$, $\gamma$, and $\alpha$. We then may write
\begin{align*}
\frac{k \p_k \, P (y)}{H (y)^2\,}
& =
\frac{k \w \p_k \, P (y)}{\left[ P (y) + \gamma \right]^2\,}
=
- k \w \p_k \, \left( \frac{1}{P (y) + \gamma} \right)
\end{align*}
and therefore, by interchanging the $k \w \p_k$- with the $y$-derivative, \begin{align*}
\frac{\dd}{\dd y} \, \frac{k \p_k \, P (y)}{H (y)^2\,}
& =
- k \w \p_k \, \frac{\dd}{\dd y} \, \left( \frac{1}{P (y) + \gamma}
\right)
=
k \w \p_k \, \frac{P' (y)}{\left[ P (y) + \gamma \right]^2\,}.
\end{align*}
As a result, eq.\ \eqref{a.18} assumes the form
\begin{align} \label{a.20}
I_1
& =
k \w \p_k \, \Bigl( k^{d-6} \, \Sigma_d ( \gamma / k^2) \Bigr)
\end{align}
with
\begin{align} \label{a.21}
\Sigma_d (\gamma/k^2)
& =
\frac{1}{d} \, k^{6-d} \, \int \limits_0^\infty \!\! \dd y~ 
y^{d/2} \, \frac{\bigl( P' (y) \bigr)^2}{\left[P(y) + \gamma \right]^4\,}.
\end{align}
Upon identifying $w \equiv \gamma / k^2$ and introducing $z \equiv y/k^2$, the function $\Sigma_d$ of eq.\ \eqref{a.21} is easily seen to coincide with the one defined by eq.\ \eqref{4.27} of the main text. Finally, in terms of the functions $\h \Sigma_d$ defined in \eqref{4.26}, we have the following result for $I_1$:
\begin{align} \label{a.22}
I_1
& =
-2 \, k^{d-6} \, \left( \gamma / k^2 \right)^{1- (6-d)/2} \,
\h \Sigma_d (\gamma / k^2).
\end{align}

In a similar way it is possible to express the second integral, $I_2$, in terms of the functions $\w \Sigma_d (w)$ defined in \eqref{4.28}. By performing the $y$-derivative in \eqref{a.19}, inserting the definitions of $P$ and $H$, and switching to dimensionless variables one is led to
\begin{align} \label{a.23}
I_2
& =
k^{d-6} \, \w \Sigma_d (\gamma / k^2).
\end{align}

If we now solve \eqref{a.17} for $\eta$ and insert \eqref{a.22} and \eqref{a.23} we arrive at our final result for the anomalous dimension:
\begin{align} \label{a.24}
\eta (k)
& =
\left( \frac{4 \, v_d \, \alpha^2}{Z_k \, k^{6-d}\,} \right) \, 
\left( \gamma / k^2 \right)^{(d-4)/2} \,
\frac{\h \Sigma_d (\gamma / k^2)}
{1 - \left( \frac{2 \, v_d \, \alpha^2}{Z_k \, k^{6-d}\,} \right) \,
\w \Sigma_d (\gamma / k^2)}.
\end{align}
Recall that $\alpha$ and $\gamma$ are given by the derivatives of $U_k$ at the reference point $\phi_1$, see eq.\ \eqref{a.9}. Notice also that $\h \Sigma_d (w)$ and $\w \Sigma_d (w)$ are dimensionless functions of a dimensionless argument.

Up to this point we were dealing with a standard scalar theory. Let us now come to the derivation of $\etan^{\text{(kin)}}$ according to the ``calculation (i)'' mentioned in Subsection \ref{4.2}. The starting point of this calculation is equation \eqref{4.9} with \eqref{4.3} inserted on its LHS. Using the notation of this appendix we have
\begin{align} \label{a.25}
\begin{split}
& k \p_k \, \left( - \frac{1}{2 \pi \, G_k} \, \frac{d-1}{d-2} \right) \, 
\int \!\! \text{d}^d x  ~
\tfrac{1}{2} \, \p_\mu \ov{f}\,(x) \, \p_\mu \ov{f}\,(x)
+ \dotsm
\\
& =
\w k\,^2 \, \tr \biggl[ 
\Big \{ \left( 1 - \etan / 2 \right) \, 
R^{(0)} ( \h p\,^2 / \w k\,^2 )
- ( \h p\,^2 / \w k\,^2 ) 
\, R^{(0) \prime} ( \h p\,^2 / \w k\,^2 )
\Big \}
\\
& \phantom{{==} \w k\,^2 \, \tr \biggl[} \times
\Bigl(\h p\,^2
+ \w k\,^2 \, R^{(0)}
(\h p\,^2 / \w k\,^2 )
- \tfrac{d (d+2)}{2 (d-1) (d-2)} \, \Lambda_k 
\bigl( \chib + \ov{f}\,(\h x\,) \bigr)^{4/(d-2)} \Bigr)^{-1}
\biggr].
\end{split}
\end{align}
Here we used that $\h R=0$ for $\h g_{\mu \nu} = \delta_{\mu \nu}$ and set 
\begin{align}\label{a.26}
\w k & \equiv \chib^{2/(d-2)} \, k.
\end{align}
Eq.\ \eqref{a.25} can be obtained from the scalar result \eqref{a.5} by replacing
\begin{gather}
\varphi \to \ov{f},
\qquad
\phi_1 \to \chib,
\qquad 
k \to \w k,
\nonumber \\ \label{a.27}
Z_k \to - \left( \frac{d-1}{2 \pi \, (d-2) \, G_k} \right),
\qquad
\eta \to \etan,
\\ \nonumber
\gamma \to - \frac{d (d+2)}{2 (d-1) (d-2)} \, \Lambda_k \,
\chib^{4/(d-2)},
\qquad
\alpha \to - \frac{2 d (d+2)}{(d-1) (d-2)^2\,} \, \Lambda_k \,
\chib^{(6-d)/(d-2)}.
\end{gather}
If we apply these substitutions to the formula \eqref{a.24} for the anomalous dimension we obtain
\begin{align} \label{a.28}
\etan (k) 
& = (-1)^{1+d/2} \, 
\frac{2^{(14-d)/2} \, \pi \, v_d \, d^{d/2} \, (d+2)^{d/2}}{(d-1)^{(d+2)/2} \, (d-2)^{(d+2)/2}\,} 
\, \,g_k \,\,\lambda_k^{d/2}
\\ \nonumber
& \phantom{{==}} \times
\h \Sigma_d \left(- \tfrac{d (d+2)}{2 (d-1) (d-2)} \, \lambda_k \right) \, 
\left[ 
1 + \frac{16 \, \pi \, v_d \, d^2 \, (d+2)^2}{(d-1)^3 \, (d-2)^3\,} \, g_k \,\,\lambda_k^2 \, \, 
\w \Sigma_d \left(- \tfrac{d (d+2)}{2 (d-1) (d-2)} \, \lambda_k \right)
\right]^{-1}
\end{align}
For $d=4$ this result reduces to eq.\ \eqref{4.25} used in the main text of the paper.

In the above derivation of $\etan$ we again observe that the renormalization properties of the CREH truncation are very different from those of a standard scalar theory. The replacement $k \to \widetilde k = \chib^{2/(d-2)} \, k$ has led to a field dependence which is non-standard. This difference will become particularly relevant when we use the more general LPA also in gravity, see \cite{creh2}.
%
%
%
\section{Generalization for $\boldsymbol{d}$ Spacetime Dimensions}\label{AppB}
In this appendix we list various key formulas which generalize the corresponding ones in the main text for an arbitrary spacetime dimensionality $d$.
The CREH truncation ansatz reads
\begin{multline}\label{b.1}
\Gamma_{k} [\,\overline{f}; \chib]
=
- \frac{d-1}{2 \pi \, (d-2) \, G_k} \,
\int \!\! \mathrm{d}^d x~\sqrt{\widehat{g}\,} \,
\Big\{ - \tfrac{1}{2} \, \left( \chib + \overline{f}\, \right)
\, \widehat{\Box} \, \left( \chib + \overline{f}\, \right)
\\
+ \tfrac{d-2}{8 (d-1)} \, \widehat{R} \, \left( \chib + \overline{f}\, \right)^2
- \tfrac{d-2}{4 (d-1)} \, \Lambda_k \, \left( \chib + \overline{f}\, \right)^{2d/(d-2)}
\Big\}.
\end{multline}
For the optimized shape function the ``kin'' version of the anomalous dimension is given by
\begin{align}
\label{b.2}
\etan^{\text{(kin)}} (g_k, \lambda_k)
& =
- \frac{32 \, \pi \, v_d \, d \, (d+2)^2}{(d-1)^3 \, (d-2)^3\,} \,\,
g_k \,\lambda_k^2
\, \left[ 1 - \frac{d \, (d+2)}{2 \, (d-1) \, (d-2)\,} \, \lambda_k \right]^{-4}.
\end{align}
Its ``pot'' version has the structure of \eqref{4.22} with the $B$-functions
\begin{align}\label{b.3}
\begin{split}
B_1 (\lambda_k)
& =
\tfrac{2}{3} \, (4 \pi)^{1-d/2} \, 
\left\{ 
\Gamma(d/2)^{-1} \,
\left[ 1 - \frac{d \, (d+2)}{2 \, (d-1) \, (d-2)\,} \, \lambda_k \right]^{-1}
\right.
\\
& \phantom{{==}\frac{2}{3} \, (4 \pi)^{1-d/2} \, \Bigg\{} \left.
- \frac{3 (d-2)}{2 (d-1)} \, \Gamma(1+d/2)^{-1} \,
\left[ 1 - \frac{d \, (d+2)}{2 \, (d-1) \, (d-2)\,} \, \lambda_k \right]^{-2}
\right\} 
\\
B_2 (\lambda_k)
& =
- \tfrac{1}{3} \, (4 \pi)^{1-d/2} \, 
\left\{ \Gamma (1+d/2)^{-1} \, 
\left[ 1 - \frac{d \, (d+2)}{2 \, (d-1) \, (d-2)\,} \, \lambda_k \right]^{-1} 
\right.
\\
& \phantom{{=} - \tfrac{1}{3} \, (4 \pi)^{1-d/2} \, \Bigg\{}
\left.
- \frac{3 (d-2)}{2 (d-1)} \, \Gamma (2+d/2)^{-1} \, \left[ 1 - \frac{d \, (d+2)}{2 \, (d-1) \, (d-2)\,} \, \lambda_k \right]^{-2} \right\}.
\end{split}
\end{align}
In $d=4$ these expressions boil down to \eqref{4.36}.
For the beta-function of the cosmological constant one obtains
\begin{align}\label{b.4}
\begin{split}
\beta_\lambda (g_k, \lambda_k)
& =
- \bigl[ 2 - \etan (g_k, \lambda_k) \bigr] \, \lambda_k
\\
& \phantom{{==}}
+ 2 \, (4 \pi)^{1-d/2} \, g_k \,
\Big\{ 
\Gamma (1+d/2)^{-1} \, 
- \tfrac{1}{2} \,
\Gamma (2+d/2)^{-1} \, \etan (g_k, \lambda_k) \Big\} \,
\\
& \phantom{{====}}
\times
\left[ 1 - \frac{d \, (d+2)}{2 \, (d-1) \, (d-2)\,} \, \lambda_k \right]^{-1}
\end{split}
\end{align}
which reduces to \eqref{4.37} for $d=4$.
These formulas were used in order to compute the entries in Table \ref{tab1}.
%
%
\pagebreak

\end{document}